\documentclass[lettersize,journal]{IEEEtran}
\usepackage{amsmath,amsfonts}
\usepackage{algorithmic}
\usepackage{array}
\usepackage[caption=false,font=normalsize,labelfont=sf,textfont=sf]{subfig}
\usepackage{textcomp}
\usepackage{stfloats}
\usepackage{url}
\usepackage{verbatim}
\usepackage{graphicx}

\usepackage{hhline}
\usepackage{rotating}
\usepackage[table]{xcolor}
\definecolor{mygray}{gray}{.93}
\usepackage{anyfontsize}

\usepackage[font=small,skip=2pt]{caption}

\usepackage{anthony}

\def\BibTeX{{\rm B\kern-.05em{\sc i\kern-.025em b}\kern-.08em
    T\kern-.1667em\lower.7ex\hbox{E}\kern-.125emX}}
\usepackage{balance}

\begin{document}
\title{Exploiting Scale-Variant Attention for \\ Segmenting Small Medical Objects}
\author{Wei Dai, \IEEEmembership{Student Member, IEEE}, Rui Liu, Zixuan Wu, Tianyi Wu, 
Min Wang, \IEEEmembership{Student Member, IEEE}, Junxian Zhou, Yixuan Yuan, \IEEEmembership{Senior Member, IEEE}, and Jun Liu, \IEEEmembership{Senior Member, IEEE}
\thanks{Wei Dai, Rui Liu, Zixuan Wu, Tianyi Wu, Min Wang, Junxian Zhou, and Jun Liu are with the Centre for Robotics and Automation, City University of Hong Kong, Hong Kong, China.
Yixuan Yuan is with the Department of Electronic Engineering, The Chinese University of Hong Kong, Hong Kong, China.
Corresponding author: jun.liu@cityu.edu.hk, yxyuan@ee.cuhk.edu.hk. 
Codes are available at: \url{https://github.com/anthonyweidai/SvANet}
}
}

\markboth{Journal of \LaTeX\ Class Files,~Vol.~18, No.~9, September~2020}%
{Exploiting Scale-Variant Attention for Segmenting Small Medical Objects}

\maketitle

\begin{abstract}
Early detection and accurate diagnosis can predict the risk of malignant disease transformation, thereby increasing the probability of effective treatment.
Identifying mild syndrome with small pathological regions serves as an ominous warning and is fundamental in the early diagnosis of diseases.
While deep learning algorithms, particularly convolutional neural networks (CNNs), have shown promise in segmenting medical objects, analyzing small areas in medical images remains challenging.
This difficulty arises due to information losses and compression defects from convolution and pooling operations in CNNs, which become more pronounced as the network deepens, especially for small medical objects.
To address these challenges, we propose a novel scale-variant attention-based network (SvANet) for accurately segmenting small-scale objects in medical images.
The SvANet consists of scale-variant attention, cross-scale guidance, Monte Carlo attention, and vision transformer, which incorporates cross-scale features and alleviates compression artifacts for enhancing the discrimination of small medical objects.
Quantitative experimental results demonstrate the superior performance of SvANet, achieving 96.12\%, 96.11\%, 89.79\%, 84.15\%, 80.25\%, 73.05\%, and 72.58\% in mean Dice coefficient for segmenting kidney tumors, skin lesions, hepatic tumors, polyps, surgical excision cells, retinal vasculatures, and sperms, which occupy less than 1\% of the image areas in KiTS23, ISIC 2018, ATLAS, PolypGen, TissueNet, FIVES, and SpermHealth datasets, respectively.
\end{abstract}

\begin{IEEEkeywords}
Small object detection, medical image segmentation, attention mechanisms, Monte Carlo method, vision transformer.
\end{IEEEkeywords}

\section{Introduction}\label{sec:intro}
\IEEEPARstart{I}{t} is essential to detect and diagnose diseases or conditions at their earliest stages, often prior to the manifestation of symptoms.
Early detection can aid in the discovery of potential diseases, such as precancerous stages, leading to higher survival rates of patients~\cite{jin2022fives, codella2019skin, ali2023multi, heller2023kits21}.
For instance, Sung \etal~\cite{sung2021global} found that the five-year survival rate for melanoma could reach 99\% if diagnosed and treated early, compared to only 27\% if detected in the late stage.
In the early stages of diseases like glaucoma~\cite{jin2022fives}, skin cancer~\cite{codella2019skin}, colorectal cancer~\cite{ali2023multi}, hepatocellular carcinoma~\cite{quinton2023tumour}, renal cancer~\cite{heller2023kits21}, \etc, the pathological areas are comparatively small and difficult to detect.
The morphometrics of these infected areas is believed to reflect the risk (\eg, cancer precursors) and progression of diseases~\cite{jin2022fives, codella2019skin, ali2023multi, quinton2023tumour, heller2023kits21, greenwald2022whole}.
Accurately delineating the boundaries of lesions is crucial for their complete resection.
Cell-level imaging analysis is also a cutting-edge field with various clinical applications, such as tumor resection analysis~\cite{greenwald2022whole} and in vitro fertilization~\cite{dai2024automated}.
However, examining cells can be challenging due to differences in size, morphology, and density, especially on a small scale.

Medical objects that occupy less than 10\% of the area pose a significant challenge in medical imaging due to their subtle texture and morphology.
Detecting the malignant potential of polyp lesions smaller than 10 mm, for example, is a challenging task~\cite{ali2023multi}. 
In reality, a large number of images from various modalities contain numerous lesions that occupy less than 10\% of the total image area~\cite{jin2022fives, codella2019skin, ali2023multi, quinton2023tumour, heller2023kits21, greenwald2022whole}, as detailed in \cref{tab:dataset}.
Deep learning algorithms, which employ convolution and pooling, can result in the loss of details for small objects, leading to noticeable compression artifacts.
Therefore, it is paramount to develop practical methods for detecting small medical objects.
To address the issue of diminished image resolution and information loss, four main strategies have been developed.
These strategies include upscaling input data~\cite{ding2021object}, expanding network variants~\cite{ronneberger2015unet, zhou2019unet++, fan2020pranet, isensee2021nnu, zhang2022hsnet, lou2023caranet, zhou2023cross, pan2023smile, dai2024deeply}, tuning loss functions~\cite{guo2018small, miao2023sc, pan2023smile}, and post-processing~\cite{pan2023smile}.

\begin{figure}[!t]
	\centering
 	\includegraphics[width=\columnwidth]{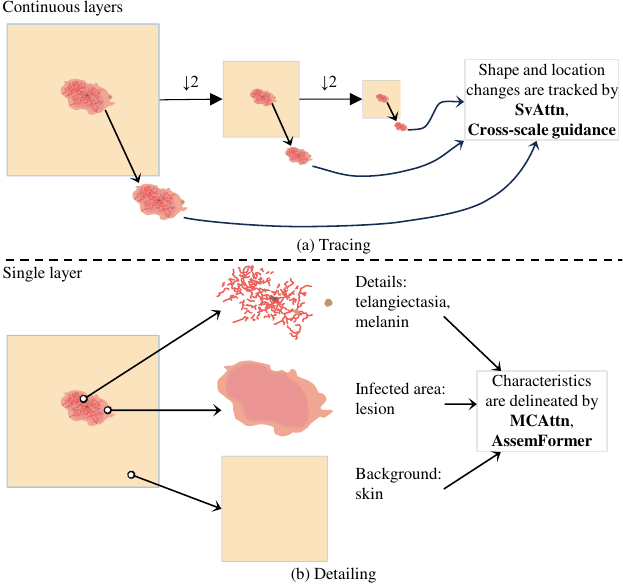}
 	\caption{
Illustration of the intuitions of the core components of the proposed methods.
The example image depicts a skin lesion.
}
 	\label{fig:intui}
\vspace{-1mm}
\end{figure}

Attention is a fundamental process for conscious perception and effective interaction with the environment.
In deep learning, the attention mechanism is an efficient method to extend network variants~\cite{fan2020pranet, lou2023caranet, zhou2023cross, pan2023smile, dai2024deeply}.
However, small medical objects pose unique challenges: they not only lack sufficient pixels and information for straightforward local representation extraction, but their relatively small size (\eg, occupying less than 1\% of the images) makes them difficult to capture using global operations such as global average pooling and multi-head self-attention.
Animal eyes provide a useful analogy for addressing these challenges by altering the shape of their crystalline lenses, thereby creating a detailed visual construction of objects and capturing changes in location and size at various distances.
Inspired by these biological mechanisms, we introduce a scale-variant attention (SvAttn) method within a cross-scale guidance module for ``\textbf{tracing}'' the behavior of small medical objects by using cross-level features, as demonstrated in \cref{fig:intui}a.

\begin{figure*}[!t]
	\centering
 	\includegraphics[width=\textwidth]{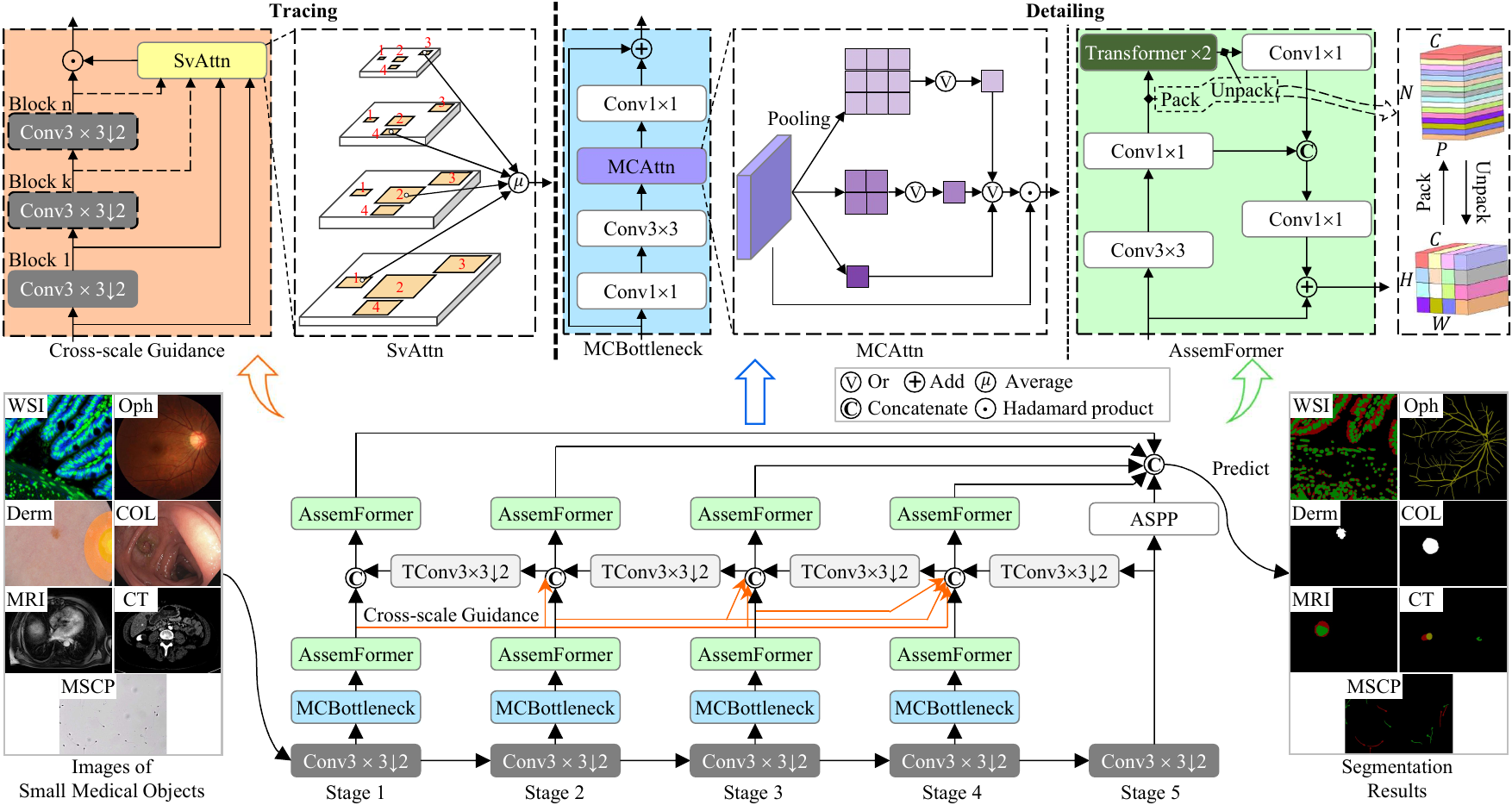}
 	\caption{
Architecture of the scale-variant attention-based network (SvANet).
Cross-scale guidance and scale-variant attention (SvAttn) techniques, depicted in the top-left dashed boxes, integrate low-level and high-level feature maps to trace the alterations in the shape and location of small medical objects.
The modules Monte Carlo attention (MCAttn) and MCAttn-based bottleneck (MCBottleneck), positioned in the top-middle dashed boxes, along with assembly-based convolutional vision transformer (AssemFormer) in the top-right dashed boxes, synergistically correlate local and global features to capture intricate object details.
}
 	\label{fig:archi}
\vspace{-1mm}
\end{figure*}

Furthermore, traditional attention mechanisms in deep learning typically produce a fixed-dimension attention map~\cite{zhou2023cross, hu2018squeeze, woo2018cbam, hou2021coordinate}, which is often inadequate to analyze medical images.
This limitation arises because these methods focus primarily on central features, neglecting the rich contextual information in the background, which is crucial for clinical interpretation.
For instance, in an abdominal slice image, the standard positional relationships among various organs (\eg, stomach, liver, kidneys, spleen, and bone marrow) aid in accurately locating objects of interest within a narrower range.
Drawing inspiration from this observation, we introduce the Monte Carlo attention (MCAttn) and an assembly-based convolutional vision transformer (AssemFormer) to address compression defects for ``\textbf{detailing}'' features of small medical objects from different scales, as illustrated by \cref{fig:intui}b.

The key contributions of this study are highlighted below:

\begin{itemize}

\item We propose SvANet, a new network that utilizes two novel attention mechanisms and a vision transformer to identify small medical objects.
To the best of our knowledge, this is the first study to systematically analyze small medical objects across seven medical image modalities and diverse object types (\ie, retinal vessels, skin lesions, polyps, livers, kidneys, tumors, tissue cells, and sperms).

\item We introduce the SvAttn method, which captures the positional and morphological essence of small medical objects by generating attention maps based on the progressively compressed feature maps.

\item We develop the MCAttn module, which generates attention maps at different scales in a single stage by using agnostic pooling output sizes.
MCAttn learns the object relations and spatial information of small medical objects with consideration of both their position and morphology.

\item We present AssemFormer, which enables the incorporation of both local spatial hierarchies and inter-patch representations, providing a comprehensive understanding of the image data.

\item Equipped with these novel designs, SvANet achieved top-level performance in ultra-small and small medical object segmentation on seven benchmark datasets, outperforming seven state-of-the-art methods.
For instance, SvANet achieved the highest mDice of 89.79\% and the lowest MAE of 1.6\texttimes10\textsuperscript{-3} in distinguishing livers and liver tumors that cover less than 1\% regions in abdominal slices.

\end{itemize}

\section{Related Work}\label{sec:relate}
\subsection{Medical Object Segmentation}\label{sec:mediseg}

Surface structures, shapes, and sizes are critical in characterizing medical objects.
The morphometric data, collected from various devices and different patients, present a complex and challenging landscape for analysis.
In recent years, deep learning algorithms have shown remarkable potential in enhancing diagnostic accuracy, reducing costs, and interpreting images of diverse medical objects across various imaging modalities.
Thes modalities include ophthalmoscopy (Oph), dermatoscopy (Derm), colonoscopy (COL), magnetic resonance imaging (MRI), computerized tomography (CT), whole slide imaging (WSI), microscopy (Microsc), electron microscopy (EM), X-ray, and \etc~\cite{ronneberger2015unet, zhou2019unet++, fan2020pranet, isensee2021nnu, zhang2022hsnet, lou2023caranet, zhou2023cross, pan2023smile, dai2024deeply}.

One widely adopted structure for analyzing medical images is the encoder-decoder-based construction, introduced by Long \etal~\cite{long2015fully} in 2015.
This approach involves extracting derived features from an encoder and using a decoder to generate the final segmentation mask.
Building upon the encoder-decoder structure, Ronneberger \etal~\cite{ronneberger2015unet} introduced ``U-shaped'' architectures, which connect the limbs by using convolution (UNet) to disseminate information for segmenting tumor cells or general objects.
To further enhance the fusion of multi-scale features in analyzing medical images across CT, MRI, and EM modalities, Zhou \etal~\cite{zhou2019unet++} introduced UNet++, an extension of UNet that incorporates densely connected links.
In addition, Isensee \etal~\cite{isensee2021nnu} broadened the application of UNet from 2D to 3D medical imaging by self-adaptive configurations (nnUNet).

To improve the performance of encoder-decoder architectures in perceiving medical images, advanced techniques have been suggested. 
These techniques consists of attention mecahnisms~\cite{jha2024transnetr}, multi network branches~\cite{pan2023smile}, contrastive learning~\cite{miao2023sc}, and feature interactions~\cite{pan2023smile, jha2024transnetr}.
For example, Fan \etal~\cite{fan2020pranet} suggested a parallel reverse attention network (PraNet) by integrating an up-sampled feature generated by the medium decoder to discern clearer boundaries of polyps in colonoscopy images.
Pan \etal~\cite{pan2023smile} introduced a three-branches ``U-shaped'' framework to ameliorate feature interactions by post-processing outputs from three branches with the watershed algorithm for examining nuclei.
In the study of CT scans of the pancreas, Miao \etal~\cite{miao2023sc} boosted the multi-branch architecture by facilitating contrastive learning and a consistency loss function. 
When assessing polyps from six unique medical centers, Jha \etal~\cite{jha2024transnetr} integrated transformer with residual connections of convolution to propagate information from the encoder to the decoder.
Despite the promising results achieved by the research above in medical image recognition, one aspect overlooked is the size of medical objects, particularly small-scale objects.

\subsection{Small Medical Object Segmentation}\label{sec:smallseg}
The convolution and pooling operations in deep learning algorithms compress input data, thus damaging the morphological characteristics of medical objects.
To mitigate information loss when reducing image resolution, one common method is to upscale the input images for generating high-resolution feature maps of small objects~\cite{ding2021object}.
However, this method can be time-consuming during training and testing due to the need for image augmentation and feature dimension enlargement.
Another promising method is to expand network variants by incorporating techniques such as atrous convolution~\cite{chen2018encoder}, skip connections~\cite{ronneberger2015unet, zhou2019unet++, isensee2021nnu}, feature pyramids~\cite{zhao2017pyramid, mei2023pyramid}, multicolumns~\cite{wang2020deep, zhang2022hsnet, zhou2023cross, pan2023smile}, or attention mechanisms~\cite{fan2020pranet, lou2023caranet, zhou2023cross, pan2023smile, dai2024deeply}, which captures cross-scale features and contributes to magnify small objects.
For example, Zhao \etal~\cite{zhao2017pyramid} introduced the pyramid scene parsing network (PSPNet), which employs pyramid pooling and concatenates up-sampled features from multiple scales to improve context feature learning.
Lou \etal~\cite{lou2023caranet} proposed a context axial reverse attention network (CaraNet) to detect small polyps and brain tumors with less than 5\% size ratios.
However, CaraNet lacks sufficient interpretability regarding its practicality for segmenting small medical objects, appearing more as a general design suited to the segmentation task.

Designing new loss functions is another practical way to boost small object identification.
Guo \etal~\cite{guo2018small} proposed a loss function that adopts the boundary pixel's neighbors to enhance the small object segmentation.
Besides, Pan \etal~\cite{pan2023smile} combined six different loss functions for nuclei diagnosis.
However, the disadvantage of replacing the loss function is that it may not be semantically understandable~\cite{guo2018small, miao2023sc} or it can increase the computational complexity~\cite{pan2023smile}.
Post-processing, such as the watershed algorithm~\cite{pan2023smile}, can also enhance small object segmentation.
However, post-processing is a distinct step from the segmentation model, and the network cannot adjust its weights to the post-processing results.

Previously, object sizes were quantified by object category~\cite{guo2018small, sang2022small}, number of pixels~\cite{ding2021object}, or size ratio\cite{lou2023caranet} in the images.
However, the size of the same object can vary based on the distance between the object and the camera, and computer vision algorithms often resize the entire input image, resulting in changes in pixel numbers.
Thus, relying solely on the object category or number of pixels cannot accurately describe the size.
This study categorizes medical objects with an area ratio below 1\% as ultra-small scale and those below 10\% as small scale, which provides a more precise and contextually relevant measure of object size tailored explicitly for medical imaging.

\subsection{Attention Mechanisms}\label{sec:attnmec}
The attention mechanism is extensively employed in semantic segmentation to prioritize salient features.
Various approaches have been proposed to incorporate attention in different ways.
Hu \etal~\cite{hu2018squeeze} applied the squeeze-excitation (SE) method to generate channel attention for learning semantic representations.
Zhou \etal~\cite{zhou2023cross} employed channel attention to capturing boundary-aware features for enhancing polyp segmentation.
To further extract spatial information, Woo \etal~\cite{woo2018cbam} combined channel attention with spatial attention in the convolution block attention module (CBAM).
Hou \etal~\cite{hou2021coordinate} further advanced CBAM by introducing coordinate attention (CoorAttn), which utilizes channel-wise average pooling to generate attention maps.
Reverse attention is another practical method to mine boundary cues.
PraNet~\cite{fan2020pranet} extracted fine-grained details by removing the estimated polyps regions using boundary information.
Lou \etal~\cite{lou2023caranet} enhanced PraNet by decomposing attention maps along height and width axes.
Relatively small feature maps (ranging from 1\texttimes1 to 44\texttimes44) were employed to bridge area and boundary cues, which may not adequately capture the structural details of minuscule objects~\cite{hu2018squeeze, fan2020pranet, dai2024deeply}. 

Moreover, self-attention is an effective attention scheme to obtain dependencies and relationships within input data.
Based on self-attention mechanisms, the vision transformer (ViT) has been introduced to process sequences of image patches to learn the inter-patch representations, which has shown noticeable potential in aggregating and preserving the features of small objects~\cite{dai2024deeply}.
He \etal~\cite{he2022fully} proposed a fully-transformer-based network that amalgamated spatial pyramid theory and vision transformer to identify skin lesions.
However, the vanilla vision transformer lacks inherent bias and is susceptible to perturbations~\cite{dosovitskiy2020image}.
Zhang \etal~\cite{zhang2022hsnet} and Pan \etal~\cite{pan2023smile} employed self-attention to improve the feature correlations in their CNN-based network for polyps and nuclei examination, respectively.
To obtain long-range information when segmenting cell nuclei, H{\"o}rst \etal~\cite{horst2024cellvit} replaced the CNN encoder with a transformer block in the UNet architecture in their CellViT model.

\section{Methodology}\label{sec:method}
\subsection{Overall Framework}\label{sec:frame}
This section introduces the scale-variant attention-based network (SvANet), specifically designed to segment small medical objects. 
The SvANet model, schematically depicted in \cref{fig:archi}, comprises four main components: cross-scale guidance in \cref{sec:csfi}, scale-variant attention in \cref{sec:svattn}, Monte Carlo attention in \cref{sec:mcattn}, and the convolution with vision transformer in \cref{sec:vit}.

\begin{figure}[!t]
	\centering
 	\includegraphics[width=\columnwidth]{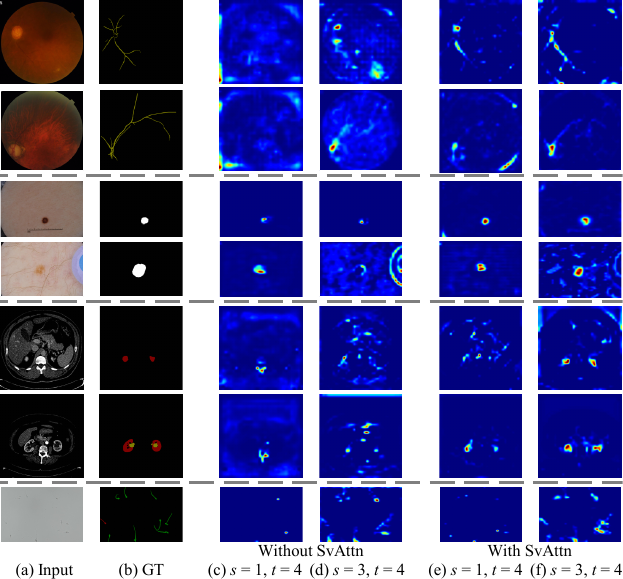}
 	\caption{
Output feature maps from cross-scale guidance without or with scale-variant attention (SvAttn).
These feature maps are generated from $g(x_s, t)$, which integrates features at different scales. 
Example images in odd and even rows include ultra-small and small medical objects, respectively (GT: ground truth).
}
 	\label{fig:svattn}
\vspace{-1mm}
\end{figure}

Preserving the features of tiny medical objects, such as sperms and retinal vessels, becomes challenging after multiple pooling or strided convolution operations.
For example, after two strided convolutions, a sperm may be reduced to being represented by only one or two pixels in the image.
In this study, cross-scale feature maps are applied to guide the latter stages in learning the features of small medical objects, as indicated by the orange arrows in \cref{fig:archi}.
The SvAttn and cross-scale guidance are primarily designed to track feature changes, particularly downsizing.
Meanwhile, MCAttn and AssemFormer distill multi-scale attention maps for improved contextual feature learning.
To better comprehend the roles of cross-scale guidance, SvAttn, MCAttn, and AssemFormer in small medical object segmentation, we examined the feature maps as shown in \cref{fig:mcattn}, \cref{fig:assemformer}, and \cref{fig:svattn}.
Due to page limitations, we selected FIVES, ISIC 2018, KiTS23, and SpermHealth datasets to visualize feature maps.
To simplicity, we chose to present the outputs from the MCBottleneck in stage four (\cref{fig:mcattn}) and two cross-scale guidance correlations (\cref{fig:svattn}).

Additionally, in \cref{fig:archi}, each Conv3\texttimes3↓2, represented by black blocks, contains a single 3\texttimes3 convolution with a stride of 2 (strided convolution).
Every TConv3\texttimes3↓2, denoted by grey blocks, consists of three convolution units: a 1\texttimes1 convolution, a 3\texttimes3 transposed strided convolution, and a 1\texttimes1 convolution.
The MCBottleneck serves as a compression point in the network, narrowing the tensor channels before expanding them to extract salient features by compressing the input information, resembling a ``bottleneck'' in information theory.
To expand the receptive field and capture features at multiple scales, atrous spatial pyramid pooling (ASPP)~\cite{chen2018encoder} is integrated after the final stage of our model.

\subsection{Cross-Scale Feature Guidance}\label{sec:csfi}
The information content decreases significantly as the size of the medical object reduces, owing to compression artifacts in neural networks.
This study introduces a cross-scale guidance module to leverage the higher-resolution features from earlier model stages.
Assume that $t$ is the target stage, the output $y_t$ can be computed as follows:
\begin{equation}\label{eq:crossscale}
y_t = \sum_{s=1}^{t - 1} g(x_s, t)
\end{equation}
where ${x_s}$ represents the input tensor in stage $s=1, 2, ..., t-1$ and the transformation $g(x_s, t)$ involves $(t-s)\ 3\times3$ strided convolutions.
The function is depicted by the orange arrows and the top-left orange blocks in \cref{fig:archi}.

As illustrated in \cref{fig:svattn}cd or ef, the highlighted region expands as the input stage increases for the same target stage, $t = 4$.
This expansion occurs due to an increased total number of strided convolution operations performed on the data.

\subsection{Scale-Variant Attention}\label{sec:svattn}
Cross-scale feature guidance is based on convolution operations, which have inherent limitations in processing global feature representations.
While global pooling operations can facilitate learning context representations, it is restricted to handling features in a uniform scale manner.
Given a subregion $x_j$ of an input tensor $x$, the output of vanilla global attention, denoted as $\mathcal{A}(x)$, is calculated as follows:
\begin{equation}\label{eq:vaattn}
\mathcal{A}(x) = \frac{1}{\sigma(x)} \sum_{j=1}^{n} x_j
\end{equation}
where $x_j$ represents the $r^2$ neighbourhood centered at the $j^{\text{th}}$ subregion of $x$. 
Here, $n$ denotes the total number of subregions, and $\sigma(x)$ represents the scalar function that normalizes the result. 
For vanilla global attention, the default values are set such that $r=1$ and $n=\sigma(x) = H_x \times W_x$.

Conventional global attention, as described in \cref{eq:vaattn}, fails to capture relationships across subregions and is limited to computing a single-scale size of the feature. 
To overcome this scale limitation while maintaining long-range correlations, we introduce scale-variant attention (SvAttn), which processes global dependencies across diverse scales, as depicted by the yellow block in \cref{fig:archi}.
In SvAttn, multi-scale attention maps are calculated across input stages $s=1, 2, ..., t-1$.
Assuming that the group-wise correspondence among input tensors is controlled by a probability $P_2(x)$, the output attention map of SvAttn is defined as:
\begin{equation}\label{eq:svattn}
\mathcal{A}_t({\bf x}) = \frac{1}{\sigma({\bf x})} \sum_{j=1}^{n} \sum_{s=1}^{t - 1} P_2(x_{s,j}) x_{s,j}
\end{equation}
where $x_{s,j}$ denotes the $j^\text{th}$ subregion of the input tensor at the $s^\text{th}$ stage, $t$ is the target stage, and ${\bf x} = [x_1, x_2, ..., x_{t-1}]^{-1}$ represents the vector of input tensors across various stages.
Besides, the correspondence probability $P_2(x)$ satisfies the conditions $\sum_{s=1}^{t - 1} P_2(x_s, j) = 1$ and $\prod_{s=1}^{t - 1} P_2(x_s, j) = 0$, thereby ensuring a weighted sum of attention maps across different scales. 
The scalar function $\sigma({\bf x})$ is defined by:
\begin{equation}
\sigma({\bf x}) = n = \frac{H \lor W}{2^{t + 1}}
\end{equation}
where H and W are the height and width of the input image, respectively.

In conjunction with \cref{eq:crossscale} and \cref{eq:svattn}, the output tensor $y_t^{'}$ of cross-scale guidance using SvAttn can be defined as follows:
\begin{equation}
y_t^{'} = \mathcal{A}_t({\bf x})y_t
\end{equation}

As indicated by \cref{eq:svattn}, the subregions located at the same proportional scaling position across stages are dynamic.
This variability enables the cross-scale guidance module to effectively discern the relationships between the high-level and low-level features.
Consequently, SvAttn enhances the network's capability to recognize downsized small medical objects throughout a sequence of stages.
As illustrated in \cref{fig:svattn}ce and df, for the same source and target stages, the features captured using SvAttn are more detailed and comprehensive for both ultra-small and small medical objects compared to those obtained without using SvAttn.
For example, from top to bottom, there is a more precise delineation of networked retinal vessels, more discernible morphology of nevi, more pronounced instance boundaries of organs such as kidneys,  and finer details in sperm morphology.
In contrast, without using SvAttn, critical features such as retinal vessels of glaucoma in the first and second rows, the nevus in the third row, and the kidneys and cyst in the sixth rows were overlooked.
It is noteworthy that ultra-small objects are harder to perceive compared to small objects without using SvAttn.
For example, moving downwards from the odd rows of  \cref{fig:svattn}cd, no retinal vessel was discovered, a relatively small nevus region was highlighted, and the left kidney was missed.

\begin{figure}[!t]
	\centering
 	\includegraphics[width=\columnwidth]{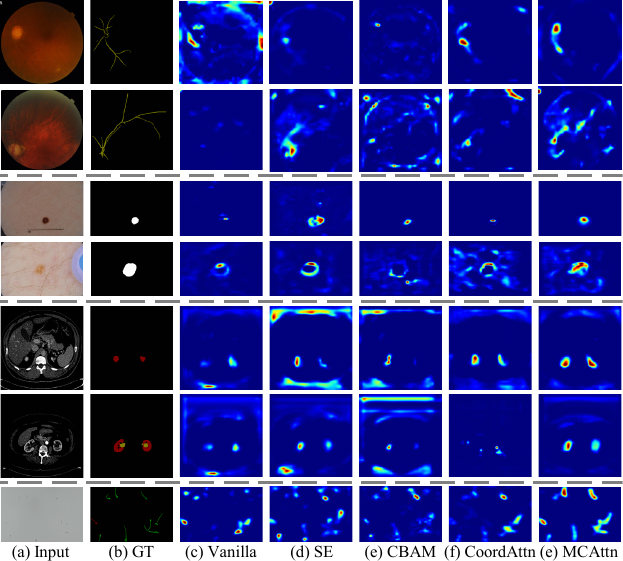}
 	\caption{
Output feature maps from the MCBottleneck without or with an attention mechanism.
Example images in odd and even rows include ultra-small and small medical objects, respectively (GT: ground truth).
}
 	\label{fig:mcattn}
\vspace{-1mm}
\end{figure}

\subsection{Monte Carlo Attention}\label{sec:mcattn}
The Monte Carlo attention (MCAttn) module, as presented by the purple block in \cref{fig:archi}, uses a random-sampling-based pooling operation to generate scale-agnostic attention maps, enabling the network to capture relevant information across different scales, enhancing its ability to identify small medical objects.
The MCAttn generates attention maps by randomly selecting a 1\texttimes1 attention map from three scales: 3\texttimes3, 2\texttimes2, and 1\texttimes1 (pooled tensors).
In conventional methods such as squeeze-excitation (SE), global-average pooling is used to acquire a 1\texttimes1 output tensor, which helps calibrate the inter-dependencies between channels~\cite{hu2018squeeze}.
However, this approach has limited capacity to exploit cross-scale correlations.
To address this limitation, MCAttn calculates the attention maps from features across three scales, thereby enhancing long-range semantic interdependences.
Given an input tensor, $x$, the output attention map of MCAttn, denoted as $\mathcal{A}_{\text{m}}(x)$, is computed as follows:
\begin{equation}\label{eq:mcattn}
\mathcal{A}_{\text{m}}(x) = \sum_{i=1}^{n} P_1(x, i) f(x, i)
\end{equation}
where $i$ denotes the output size of the attention map, and $f(x, i)$ represents the average pooling function. 
The association probability $P_1(x, i)$ satisfies the conditions $ \sum_{i=1}^{n} P_1(x, i) = 1$ and $\prod_{i=1}^{n} P_1(x, i) = 0$, ensuring the generation of agnostic and generalizable attention maps.
$n$ represents the number of output pooled tensors and is set to 3 in this study.

The Monte Carlo sampling method described in \cref{eq:mcattn} allows for the random selection of association probabilities, enabling the extraction of both local information (\eg, angle, edge, and color) and context information (\eg, whole image texture, spatial correlation, and color distribution).
In \cref{fig:mcattn}c, the second to the fourth rows and the final row illustrate that MCBottleneck, without using an attention mechanism, struggles to detect the retinas and nevi and often overlooks several sperms.
Conversely, when attention mechanisms like SE, CBAM, and CoordAttn are used, localization of densely occupied regions (\eg, optic disc, kidneys, and sperms)  is enhanced compared to when no attention mechanism is used.
However, sparse regions, such as retinal vasculatures and nevus centers, are often underlooked, especially the ultra-small ones, as shown in \cref{fig:mcattn}c-f.
Instead, using MCAttn in MCBottleneck, as depicted in \cref{fig:mcattn}cg, enhances the discernibility of the morphology and precise location of both ultra-small and small medical objects compared to when MCAttn is not used.
For instance, in \cref{fig:mcattn}, moving downwards, MCBottleneck coupled with MCAttn emphasizes more apparent retinal vessels for glaucoma, sharper boundaries of nevi, and more perceptible morphology of kidneys, cysts, and sperms.
MCAttn also accentuates other medical objects of interest, such as retinas, nevi, kidneys, and sperms, as shown in \cref{fig:mcattn}g.

\subsection{Convolution with Vision Transformer}\label{sec:vit}
The proposed assembly-based convolutional vision transformer, termed AssemFormer, is illustrated in top-right dased green boxes in \cref{fig:archi}.
Inspired by~\cite{mehta2021mobilevit, dai2024deeply}, AssemFormer incorporates a 3\texttimes3 convolution and a 1\texttimes1 convolution, followed by two transformer blocks and two convolution operations.
AssemFormer bridges convolution and transformer operations by stacking and unstacking feature maps.
Equipped with this design, AssemFormer tackles the lack of inductive biases for the vanilla transformer.

\begin{figure}[!t]
	\centering
 	\includegraphics[width=\columnwidth]{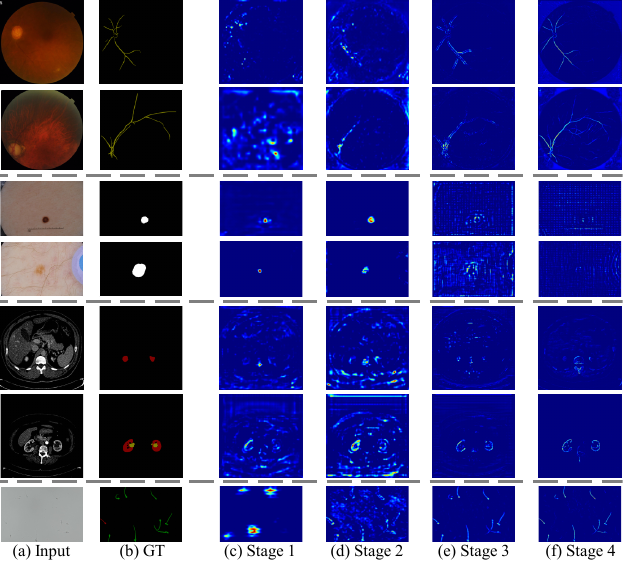}
 	\caption{
Output feature maps from the AssemFormer. 
The feature maps are extracted from four encoder stages individually (GT: ground truth).
Since the layer in the fifth stage is directly connected with the ASPP module, no AssemFormer is used at this stage.
}
 	\label{fig:assemformer}
\vspace{-1mm}
\end{figure}

The functionalities of convolution and transformer operations differ.
Convolution operations focus on learning local and general features, such as corners, edges, angles, and colors of medical objects.
In contrast, the transformer module extracts global information, including morphology, depth, and color distribution of medical objects, utilizing multi-head self-attention (MHSA).
In addition, the transformer module also learns positional associations of medical objects, such as the relationships between a tumor and the kidney, a kidney and the abdomen, and a tumor and the abdomen within an MRI slice image.
The vision transformer algorithm employs a sequence of MHSA and multilayer perceptron (MLP) blocks, each followed by layer normalization~\cite{dosovitskiy2020image}.
The self-attention mechanism~\cite{vaswani2017attention} is formulated as follows:
\begin{equation}\label{eq:assemformer}
\mathcal{A}_{\text{ViT}}(\bold{q}, \bold{k}, \bold{v}) = \text{softmax}(\frac{\bold{q}\bold{k}^T}{\sqrt{D_\text{h}}})\bold{v}
\end{equation}
where $\bold{q}$, $\bold{k}$, and $\bold{v}$ are the query, key, and value vectors of an input sequence $\bold{z} \in \mathbb{R}^{N\times D}$. $N$ denotes the number of patches, and $D$ represents the patch size. Given $m$ self-attention operations, $D_\text{h}$, the dimension of $\bold{q}$ and $\bold{k}$, is defined as $D / m$.

Furthermore, a skip connection and concatenation are incorporated to mitigate the information loss concerning small medical objects.
Leveraging the convolution-transformer hybrid structure, the AssemFormer block can simultaneously learn the local and global representation of an input medical image.
According to the ablation study presented in the fifth and sixth rows of \cref{tab:ablation}, the AssemFormer significantly improves the segmentation performance of SvANet.

In \cref{fig:assemformer}, progressing from left to right, the AssemFormer increasingly highlights smaller areas that more accurately align with the ground truth, especially notable in scenarios with fewer medical objects.
For instance, the first row of \cref{fig:mcattn} demonstrates how the thin lines of retinal vasculature and light reflections are initially emphasized, becoming progressively thicker.
Subsequently, these lines become shorter and more focused on a smaller region corresponding to the optic disc location, as depicted in the second first of \cref{fig:mcattn}ef.
Large-scale distortions, such as noise or compression defects, play a role in this observed trend where the concentration of feature maps intensifies with a deeper layer.
The pattern of increased feature map concentration is consistent across the segmentation of various medical objects, including skin lesions, polyps, hepatic tumors, livers, kidneys, tissue cells, and sperms.

The multi-head self-attention mechanism of AssemFormer, described in \cref{eq:assemformer}, facilitates patch interactions and enriches the context information.
In contrast to \cref{fig:mcattn}, from left to right, the feature maps evolve from AssemFormer from coarse to fine representations.
As illustrated in \cref{fig:mcattn}f, the AssemFormer enhances the visibility and precise localization of small medical objects such as glaucoma, nevus, polyp, hepatic tumor, kidney, tissue cell, and sperm, highlighting their morphological details and exact positions.

\section{Experimental Results}\label{sec:exp}
\subsection{Evaluation Protocol}\label{sec:eval}
\subsubsection{Dataset}\label{sec:dataset}
To validate the effectiveness of SvANet, we conducted tests alongside seven state-of-the-art (SOTA) models for small medical object segmentation across seven benchmark datasets: FIVES~\cite{jin2022fives}, ISIC 2018~\cite{codella2019skin}, PolypGen~\cite{ali2023multi}, ATLAS~\cite{quinton2023tumour}, KiTS23~\cite{heller2023kits21}, TissueNet~\cite{greenwald2022whole}, and SpermHealth.

The FIVES dataset comprises 800 fundus photographs taken with ophthalmoscopes featuring age-related macular degeneration, diabetic retinopathy, glaucoma, and health fundus types.
The ISIC 2018 dataset includes skin lesion images collected by dermatoscopes, encompassing healthy and unhealthy skin areas.
PolypGen, sourced from six different hospitals using colonoscopes, focuses exclusively on polyps.
The ATLAS dataset consists of 90 MRI scans of livers, detailing two types of medical objects: the liver and tumor.
KiTS23 has 599 CT scans of kidneys, categorized into three semantic classes (\ie, kidney, tumor, and cyst).
For experimental comparisons, the ATLAS and KiTS23 datasets were sliced into sequences of 2-D images.
In addition, TissueNet includes images of cells from the pancreas, breast, tonsil, colon, lymph, lung, esophagus, skin, and spleen, derived from humans, mice, and macaques, collected using cell imaging platforms such as CODEX and CyCIF, with annotations for whole cells and their nuclei.

\begin{table}[!t]
	\centering
	\scriptsize
	\renewcommand{\arraystretch}{1.1}
	\setlength\tabcolsep{5pt}
	\caption{
Dataset details: medical objects within each dataset are categorized by area ratios: below 1\% (ultra-small), below 10\% (small), and below 100\% (all).
}
	\label{tab:dataset}
\resizebox{\columnwidth}{!}{
	\begin{tabular}{l|*{3}{c|}*{4}{c}}
	\Xhline{3\arrayrulewidth}
	\rowcolor{mygray}
	& & Number of & & \multicolumn{3}{c}{Number of object} \\ [-0.5pt] \hhline{*{4}{~}|*{4}{-}}
	\rowcolor{mygray}
	& \multirow{-2}{*}{Image} & image & & ultra & & \\
	\rowcolor{mygray}	
	\multirow{-3}{*}{Dataset} & \multirow{-2}{*}{capture} & (train + test) & \multirow{-3}{*}{Object area ratio} & small & \multirow{-2}{*}{small} & \multirow{-2}{*}{all} \\
	\Xhline{1.5\arrayrulewidth}

	FIVES~\cite{jin2022fives} & Oph & 600 + 200 & 0.351\% $\sim$ 52.020\% & 4 & 145 & 798 \\ \hline

	ISIC 2018~\cite{codella2019skin} & Derm & 2594 + 100 & 0.288\% $\sim$ 98.575\% & 52 & 1084 & 2694 \\ \hline

	PolypGen~\cite{ali2023multi} & COL & 1230 + 307 & 0.003\% $\sim$ 85.850\% & 81 & 895 & 1411 \\ \hline

	ATLAS~\cite{quinton2023tumour} & MRI & 997 + 249 & 0.001\% $\sim$ 25.826\% & 274 & 1084 & 1464 \\ \hline

	KiTS23~\cite{heller2023kits21} & CT & 1703 + 426 & 0.001\% $\sim$ 13.790\% & 665 & 1533 & 1539 \\ \hline

	TissueNet~\cite{greenwald2022whole} & WSI  & 2580 + 1324 & 0.002\% $\sim$ 9.836\% & 9096 & 9437 & 9437 \\ \hline

	SpermHealth & Microsc & 118 + 30 & 0.042\% $\sim$ 0.651\% & 1456 & 1456 & 1456 \\ 

	\Xhline{3\arrayrulewidth}
	\end{tabular}
}
\vspace{-1mm}
\end{table}

SpermHealth is a customized dataset from the 3rd Affiliated Hospital of Shenzhen University, consisting of low-resolution sperm images (640\texttimes480, 96 DPI) extracted from microscope-captured videos.
These images have been meticulously annotated into normal and abnormal categories by experienced fertility doctors.
Further details of the datasets used in the tests are presented in \cref{tab:dataset}.


\subsubsection{Implementation Details and Evaluation Metric}\label{sec:imple}
In this study, the mini-batch size was set to 4. 
Data augmentation strategies applied to pre-process the input images included random horizontal flips, random cropping to a resolution of 512\texttimes512, Gaussian blur, distortion, and rotation. 
The AdamW optimizer~\cite{loshchilov2017decoupled} and a cross-entropy loss function were utilized, with the learning rate decaying from 5\texttimes10\textsuperscript{-5} to 1\texttimes10\textsuperscript{-6} following a cosine schedule~\cite{loshchilov2016sgdr}.
The total training process spanned 100 epochs. 
The results were calculated by averaging the outcomes from three times of training and testing cycles.
All backbone was pretrained in the ImageNet-1K~\cite{deng2009imagenet} dataset.
Additionally, all tested methods followed the configurations above of training except that nnUNet utilized official settings~\cite{isensee2021nnu} for training.

The experiments were conducted on an RTX 4090 GPU with an AMD Ryzen 9 7950X CPU.
The metrics used to assess the performance of semantic segmentation include the mean Dice coefficient (mDice),  mean intersection over union (mIoU), mean absolute error (MAE), sensitivity, and F2 score.

\subsection{Results for Datasets with Diverse Object Sizes}\label{sec:results1}
The experimental results for the FIVES, ISIC 2018, PolypGen, ATLAS, KiTS23, and TissueNet datasets are summarized in \cref{tab:segresults}.
These results demonstrate that SvANet outperforms other SOTA methods across all metrics for ultra-small and small medical object segmentation across six datasets tested.

As presented in \cref{tab:segresults}, SvANet outperformed other SOTA methods across three object scales in the FIVES, ISIC 2018, and ATLAS datasets, excluding sensitivity of 93.54\% \& 87.13\% in ISIC 2018 \& ATLAS datasets and MAE of 5.35\texttimes10\textsuperscript{-4} \& 6.6\texttimes10\textsuperscript{-3} in FIVES \& ATLAS datasets for the ``all'' object scale, as summarized in \cref{tab:segresults}.
Additionally, SvANet surpasses other methods with increments in the mDice of at least +2.95\% \& +5.23\%, mIoU of +1.97\% \& +5.78\%, sensitivity of +0.19\% \& +5.03\%, and F2 score of +1.28\% \& +5.15\% for differentiating ultra-small and small retinal vessels in the FIVES dataset.
However, the MAE is comparatively high ($>$7.5\texttimes10\textsuperscript{-3}) in ultra-small retinal vasculature segmentation across all tested models, potentially due to the minimal number (4) ultra-small objects providing insufficient learnable features for deep learning algorithms.
In ISIC 2018 and ATLAS datasets, SvANet excelled in segmenting ultra-small objects (\ie, skin lesions, livers, and hepatic tumors) with mDice of 96.11\% \& 89.79\%, mIoU of 92.76\% \& 86.06\%, sensitivity of 98.35\% \& 86.68\%, and F2 score of 97.42\% \& 87.71\%.
These results suggest significant potential for SvANet in diagnosing dermatological skin lesions and hepatic tumors in MRI scans, particularly for objects with an area ratio smaller than 1\% or 10\%.
Thus, SvANet can ameliorate therapeutic approaches such as excision therapy, laser therapy, electrosurgery, and radiotherapy for treating these conditions.

Furthermore, the segmentation results for the PolypGen and KiTS23 datasets demonstrate that SvANet delivers superior performance than other SOTA methods across three object scales.
Specifically, SvANet achieved the highest mDice of 84.15\% \& 96.12\%, 91.17\% \& 94.01\%, and 93.16\% \& 94.54\% for ultra-small, small, and all medical object scales in PolypGen and KiTS23 datasets, respectively.
Moreover, SvANet delivered up to 14.83\% \& 2.76\%, 6.23\% \& 6.88\%, and 6.93\% \& 6.33\% increments in F2 score over other tested methods for ultra-small, small, and all object scales in PolypGen and KiTS23 datasets, separately.
The F2 score, the harmonic mean of sensitivity and precision, underscores the robustness of SvANet in medical object segmentation.
SvANet also recorded the lowest MAE, 1.01\texttimes10\textsuperscript{-4} \& 2.0\texttimes10\textsuperscript{-2}, 6.6\texttimes10\textsuperscript{-3} \& 7.0\texttimes10\textsuperscript{-2}, and 8.1\texttimes10\textsuperscript{-3} \& 8.0\texttimes10\textsuperscript{-2} across three object scales for PolypGen and KiTS23 datasets, indicating a high level of precision in the pixel-level recognition of polyps, kidneys, renal tumors, and cysts.

\begin{table*}[!t]
	\centering
	\scriptsize
	\renewcommand{\arraystretch}{1.1}
	\setlength\tabcolsep{5pt}
	\caption{
Quantitative results in FIVES, ISIC 2018, PolypGen, ATLAS, KITS23, and TissueNet datasets, divided by area ratios of medical objects: below 1\% (ultra-small), below 10\% (small), and below 100\% (all). 
The best results are underlined in bold.
}\label{tab:segresults}
	\begin{tabular}{cr||*{3}{c}|*{3}{c}|*{3}{c}|*{3}{c}|*{3}{c}}

	\Xhline{4\arrayrulewidth}
	\rowcolor{mygray}
	& & \multicolumn{3}{c|}{mDice} & \multicolumn{3}{c|}{mIoU} & \multicolumn{3}{c|}{MAE (\texttimes10\textsuperscript{-4})} & \multicolumn{3}{c|}{Sensitivity} & \multicolumn{3}{c}{F2 score} \\ [-0.5pt] \hhline{~~|*{15}{-}}
	\rowcolor{mygray}
	&	& ultra &	&	& ultra &	&	& ultra &	&	& ultra &	&	& ultra &	& \\
	\rowcolor{mygray}	
	& \multirow{-3}{*}{Methods} & small & \multirow{-2}{*}{small} & \multirow{-2}{*}{all} & small & \multirow{-2}{*}{small} & \multirow{-2}{*}{all} & small & \multirow{-2}{*}{small} & \multirow{-2}{*}{all} & small & \multirow{-2}{*}{small} & \multirow{-2}{*}{all} & small & \multirow{-2}{*}{small} & \multirow{-2}{*}{all}\\
	\Xhline{2\arrayrulewidth}

	\multirow{8}{*}{\begin{sideways}FIVES\end{sideways}}
	& UNet (MICCAI'15)~\cite{ronneberger2015unet} & 67.49 & 72.99 & 71.12 & 62.54 & 65.04 & 63.10 & 84.65 & 14.64 & 6.22 & 63.86 & 71.48 & 70.13 & 65.06 & 72.02 & 70.42 \\
	& UNet++ (TMI'19)~\cite{zhou2019unet++} & 70.10 & 74.46 & 70.72 & 65.06 & 66.37 & 62.68 & 80.64 & 14.46 & 5.92 & 67.02 & 73.05 & 69.50 & 68.08 & 73.58 & 69.90 \\
	& HRNet (TPAMI'20)~\cite{wang2020deep} & 69.46 & 79.16 & 75.66 & 64.67 & 70.83 & 67.04 & 88.84 & 15.38 & 5.51 & 68.33 & 78.30 & 74.09 & 68.68 & 78.60 & 74.67 \\
	& PraNet (MICCAI'20)~\cite{fan2020pranet} & 64.47 & 71.46 & 68.57 & 59.42 & 63.08 & 60.14 & 79.64 & 14.78 & 5.45 & 61.20 & 67.88 & 65.61 & 62.27 & 69.11 & 66.58 \\
	& nnUNet (NM'21)~\cite{isensee2021nnu} & 58.02 & 74.34 & 71.73 & 54.16 & 66.27 & 63.59 & 91.39 & 14.70 & 5.90 & 54.80 & 73.39 & 71.03 & 55.68 & 73.72 & 71.27 \\
	& CFANet (PR'23)~\cite{zhou2023cross} & 69.35 & 76.64 & 72.23 & 63.16 & 68.60 & 65.27 & 80.18 & 15.16 & \textbf{4.92} & 64.86 & 75.67 & 73.79 & 66.29 & 75.99 & 72.75 \\
	& TransNetR (MIDL'24)~\cite{jha2024transnetr} & 67.87 & 80.68 & 78.60 & 63.59 & 72.55 & 69.66 & 83.54 & 14.44 & 5.68 & 66.90 & 78.83 & 77.24 & 67.26 & 79.45 & 77.60 \\
	& \textbf{SvANet~(Ours)} & \textbf{73.05} & \textbf{85.91} & \textbf{86.29} & \textbf{67.03} & \textbf{78.33} & \textbf{78.39} & \textbf{76.82} & \textbf{14.42} & 5.35 & \textbf{68.52} & \textbf{83.86} & \textbf{85.15} & \textbf{69.96} & \textbf{84.60} & \textbf{85.46} \\
	\hline
	\hline

	\multirow{8}{*}{\begin{sideways}ISIC 2018\end{sideways}}
	& UNet (MICCAI'15)~\cite{ronneberger2015unet} & 88.00 & 89.77 & 90.87 & 80.98 & 82.59 & 83.55 & 47.51 & 11.22 & 7.13 & 98.23 & 93.67 & 90.61 & 93.35 & 92.00 & 90.71 \\
	& UNet++ (TMI'19)~\cite{zhou2019unet++} & 81.10 & 88.80 & 90.73 & 72.61 & 81.20 & 83.30 & 93.65 & 12.59 & 7.35 & 98.29 & 93.64 & 91.06 & 89.12 & 91.54 & 90.92 \\
	& HRNet (TPAMI'20)~\cite{wang2020deep} & 88.32 & 89.83 & 91.84 & 81.11 & 82.67 & 85.13 & 43.62 & 11.53 & 6.44 & 97.94 & 95.23 & 92.02 & 93.47 & 92.87 & 91.95 \\
	& PraNet (MICCAI'20)~\cite{fan2020pranet} & 95.04 & 90.66 & 93.03 & 90.96 & 83.90 & 87.14 & 15.01 & 10.46 & 5.55 & 97.01 & 95.56 & \textbf{93.64} & 96.16 & 93.44 & 93.39 \\
	& nnUNet (NM'21)~\cite{isensee2021nnu} & 89.29 & 89.74 & 91.01 & 82.90 & 82.60 & 83.78 & 41.96 & 11.47 & 7.02 & 98.35 & 94.19 & 90.74 & 94.02 & 92.26 & 90.84 \\
	& CFANet (PR'23)~\cite{zhou2023cross} & 94.09 & 90.34 & 92.89 & 89.51 & 83.41 & 86.89 & 18.64 & 10.93 & 5.63 & 97.08 & 95.68 & 93.23 & 95.82 & 93.35 & 93.09 \\
	& TransNetR (MIDL'24)~\cite{jha2024transnetr} & 88.73 & 90.43 & 92.35 & 82.07 & 83.56 & 85.99 & 42.82 & 10.69 & 6.01 & 96.67 & 95.21 & 92.38 & 92.92 & 93.14 & 92.36 \\
	& \textbf{SvANet~(Ours)} & \textbf{96.11} & \textbf{91.63} & \textbf{93.24} & \textbf{92.76} & \textbf{85.36} & \textbf{87.50} & \textbf{11.90} & \textbf{9.18} & \textbf{5.35} & \textbf{98.35} & \textbf{95.71} & 93.54 & \textbf{97.42} & \textbf{93.96} & \textbf{93.42} \\
	\hline
	\hline

	\multirow{8}{*}{\begin{sideways}PolypGen\end{sideways}}
	& UNet (MICCAI'15)~\cite{ronneberger2015unet} & 73.40 & 84.81 & 87.14 & 70.96 & 76.50 & 78.85 & 2.94 & 1.13 & 1.45 & 79.18 & 84.09 & 84.06 & 75.58 & 84.36 & 85.19 \\
	& UNet++ (TMI'19)~\cite{zhou2019unet++} & 74.87 & 85.79 & 88.43 & 71.86 & 77.67 & 80.60 & 2.81 & 1.07 & 1.33 & 82.84 & 85.44 & 85.85 & 77.89 & 85.58 & 86.82 \\
	& HRNet (TPAMI'20)~\cite{wang2020deep} & 70.58 & 85.34 & 89.32 & 68.88 & 77.10 & 81.85 & 5.72 & 1.14 & 1.26 & 76.36 & 85.77 & 87.38 & 71.93 & 85.58 & 88.12 \\
	& PraNet (MICCAI'20)~\cite{fan2020pranet} & 81.11 & 90.69 & 92.60 & 76.41 & 84.22 & 86.83 & 1.34 & 0.68 & 0.88 & 86.99 & 89.03 & 90.71 & 83.97 & 89.67 & 91.44 \\
	& nnUNet (NM'21)~\cite{isensee2021nnu} & 78.52 & 87.94 & 89.33 & 77.50 & 80.41 & 81.87 & 4.06 & 0.94 & 1.24 & 84.95 & 88.52 & 87.04 & 79.52 & 88.29 & 87.91 \\
	& CFANet (PR'23)~\cite{zhou2023cross} & 79.44 & 90.65 & 92.71 & 75.08 & 84.16 & 87.00 & 1.75 & 0.70 & 0.86 & 87.76 & 88.79 & 90.70 & 83.16 & 89.51 & 91.47 \\
	& TransNetR (MIDL'24)~\cite{jha2024transnetr} & 79.51 & 90.67 & 92.49 & 75.16 & 84.19 & 86.65 & 1.80 & 0.70 & 0.89 & 88.08 & 90.04 & 90.68 & 83.29 & 90.29 & 91.37 \\
	& \textbf{SvANet~(Ours)} & \textbf{84.15} & \textbf{91.17} & \textbf{93.16} & \textbf{78.95} & \textbf{84.90} & \textbf{87.71} & \textbf{1.01} & \textbf{0.66} & \textbf{0.81} & \textbf{89.21} & \textbf{90.21} & \textbf{91.47} & \textbf{86.76} & \textbf{90.59} & \textbf{92.12} \\
	\hline
	\hline

	\multirow{8}{*}{\begin{sideways}ATLAS\end{sideways}} 
	& UNet (MICCAI'15)~\cite{ronneberger2015unet} & 82.09 & 83.55 & 85.45 & 79.89 & 76.60 & 77.89 & 0.41 & 0.75 & 0.88 & 81.98 & 81.24 & 83.40 & 81.95 & 82.05 & 84.12 \\
	& UNet++ (TMI'19)~\cite{zhou2019unet++} & 81.70 & 83.91 & 84.75 & 79.58 & 76.98 & 77.33 & 0.47 & 0.73 & 0.84 & 82.59 & 82.33 & 83.00 & 82.17 & 82.86 & 83.57 \\
	& HRNet (TPAMI'20)~\cite{wang2020deep} & 85.86 & 84.98 & 86.66 & 82.56 & 78.53 & 79.68 & 0.26 & 0.56 & \textbf{0.65} & 84.74 & 83.70 & 85.12 & 85.14 & 84.09 & 85.59 \\
	& PraNet (MICCAI'20)~\cite{fan2020pranet} & 86.04 & 86.70 & 88.02 & 82.69 & 80.00 & 81.12 & 0.30 & 0.63 & 0.76 & 85.39 & 85.22 & 86.80 & 85.62 & 85.77 & 87.25 \\
	& nnUNet (NM'21)~\cite{isensee2021nnu} & 86.22 & 85.01 & 85.65 & 85.79 & 77.93 & 77.97 & 0.23 & 0.82 & 1.01 & 86.46 & 83.74 & 84.83 & 86.35 & 84.19 & 85.13 \\
	& CFANet (PR'23)~\cite{zhou2023cross} & 86.24 & 87.04 & 88.25 & 82.96 & 80.52 & 81.46 & 0.26 & 0.57 & 0.70 & 84.89 & 85.51 & 86.44 & 85.34 & 86.03 & 87.06 \\
	& TransNetR (MIDL'24)~\cite{jha2024transnetr} & 86.28 & 86.53 & 88.69 & 82.93 & 80.37 &  82.05 & 0.27 & 0.51 & 0.66 & 85.29 & 85.42 & \textbf{87.15} & 85.61 & 85.75 & 87.68 \\
	& \textbf{SvANet~(Ours)} & \textbf{89.79} & \textbf{87.60} & \textbf{89.14} & \textbf{86.06} & \textbf{81.29} & \textbf{82.56} & \textbf{0.16} & \textbf{0.51} & 0.66 & \textbf{86.68} & \textbf{85.86} & 87.13 & \textbf{87.71} & \textbf{86.43} & \textbf{87.82} \\
	\hline
	\hline

	\multirow{8}{*}{\begin{sideways}KiTS23\end{sideways}}
	& UNet (MICCAI'15)~\cite{ronneberger2015unet} & 95.28 & 91.27 & 91.94 & 93.42 & 87.58 & 88.30 & 0.03 & 0.07 & 0.09 & 96.21 & 91.08 & 91.54 & 95.76 & 91.11 & 91.65 \\
	& UNet++ (TMI'19)~\cite{zhou2019unet++} & 95.38 & 93.94 & 93.73 & 93.50 & 89.14 & 89.86 & 0.04 & 0.08 & 0.09 & 96.40 & 92.79 & 93.28 & 95.90 & 92.89 & 93.45 \\
	& HRNet (TPAMI'20)~\cite{wang2020deep} & 96.00 & 93.48 & 93.98 & 94.25 & 89.72 & 90.35 & 0.02 & 0.07 & 0.09 & 96.16 & 93.08 & 93.52 & 96.09 & 93.21 & 93.67 \\
	& PraNet (MICCAI'20)~\cite{fan2020pranet} & 95.58 & 92.84 & 93.39 & 93.67 & 88.71 & 89.34 & 0.03 & 0.08 & 0.10 & 96.25 & 92.38 & 92.84 & 95.95 & 92.53 & 93.03 \\
	& nnUNet (NM'21)~\cite{isensee2021nnu} & 93.05 & 87.26 & 88.46 & 91.12 & 83.77 & 84.68 & 0.03 & 0.07 & 0.09 & 94.26 & 86.76 & 87.73 & 93.74 & 86.92 & 87.98 \\
	& CFANet (PR'23)~\cite{zhou2023cross} & 95.33 & 93.66 & 94.14 & 93.57 & 89.80 & 90.39 & 0.02 & 0.07 & 0.09 & 96.10 & 93.15 & 93.60 & 95.76 & 93.32 & 93.78 \\
	& TransNetR (MIDL'24)~\cite{jha2024transnetr} & 95.82 & 93.12 & 93.76 & 94.20 & 89.51 & 90.31 & 0.02 & 0.07 & 0.08 & 96.65 & 92.79 & 93.42 & 96.30 & 92.88 & 93.52 \\
	& \textbf{SvANet~(Ours)} & \textbf{96.12} & \textbf{94.01} & \textbf{94.54} & \textbf{94.51} & \textbf{90.38} & \textbf{91.05} & \textbf{0.02} & \textbf{0.07} & \textbf{0.08} & \textbf{96.81} & \textbf{93.70} & \textbf{94.20} & \textbf{96.50} & \textbf{93.80} & \textbf{94.31} \\
	\hline
	\hline

	\multirow{8}{*}{\begin{sideways}TissueNet\end{sideways}}
	& UNet (MICCAI'15)~\cite{ronneberger2015unet} & 65.89 & \multicolumn{2}{c|}{86.36} & 56.03 & \multicolumn{2}{c|}{76.99} & 28.26 & \multicolumn{2}{c|}{3.34} & 80.69 & \multicolumn{2}{c|}{86.29} & 71.29 & \multicolumn{2}{c}{86.32} \\
	& UNet++ (TMI'19)~\cite{zhou2019unet++} & 64.43 & \multicolumn{2}{c|}{85.89} & 52.73 & \multicolumn{2}{c|}{76.25} & 41.76 & \multicolumn{2}{c|}{3.39} & 78.36 & \multicolumn{2}{c|}{85.89} & 67.19 & \multicolumn{2}{c}{85.88} \\
	& HRNet (TPAMI'20)~\cite{wang2020deep} & 64.66 & \multicolumn{2}{c|}{86.99} & 54.96 & \multicolumn{2}{c|}{77.84} & 31.61 & \multicolumn{2}{c|}{3.35} & 78.18 & \multicolumn{2}{c|}{86.92} & 69.15 & \multicolumn{2}{c}{86.94} \\
	& PraNet (MICCAI'20)~\cite{fan2020pranet} & 60.30 & \multicolumn{2}{c|}{85.96} & 50.81 & \multicolumn{2}{c|}{76.36} & 47.43 & \multicolumn{2}{c|}{3.37} & 77.41 & \multicolumn{2}{c|}{85.98} & 64.84 & \multicolumn{2}{c}{85.97} \\
	& nnUNet (NM'21)~\cite{isensee2021nnu} & 61.35 & \multicolumn{2}{c|}{86.65} & 55.36 & \multicolumn{2}{c|}{77.38} & 45.41 & \multicolumn{2}{c|}{3.33} & 82.53 & \multicolumn{2}{c|}{86.74} & 66.39 & \multicolumn{2}{c}{86.70} \\
	& CFANet (PR'23)~\cite{zhou2023cross} & 71.75 & \multicolumn{2}{c|}{87.48} & 62.00 & \multicolumn{2}{c|}{78.59} & 16.15 & \multicolumn{2}{c|}{3.31} & 80.20 & \multicolumn{2}{c|}{87.43} & 75.39 & \multicolumn{2}{c}{87.45} \\
	& TransNetR (MIDL'24)~\cite{jha2024transnetr} & 65.74 & \multicolumn{2}{c|}{86.85} & 56.07 & \multicolumn{2}{c|}{77.69} & 24.28 & \multicolumn{2}{c|}{3.34} & 61.70 & \multicolumn{2}{c|}{86.94} & 71.46 & \multicolumn{2}{c}{86.90} \\
	& \textbf{SvANet~(Ours)} & \textbf{80.25} & \multicolumn{2}{c|}{\textbf{88.05}} & \textbf{71.60} & \multicolumn{2}{c|}{\textbf{79.45}} & \textbf{7.22} & \multicolumn{2}{c|}{\textbf{3.28}} & \textbf{83.36} & \multicolumn{2}{c|}{\textbf{88.07}} & \textbf{82.00} & \multicolumn{2}{c}{\textbf{88.06}} \\

	\Xhline{4\arrayrulewidth}
	\end{tabular}
\vspace{-4mm} 
\end{table*}

In the TissueNet dataset, which includes only ultra-small and small cells, \cref{tab:segresults} reveals that the SvANet leads in segmentation performance, achieving 80.25\% \& 88.05\% in mDice, 71.60\% \& 79.45\% in mIoU, 7.22\texttimes10\textsuperscript{-4} \& 3.28\texttimes10\textsuperscript{-4} in MAE, 83.36\% \& 88.07\% in sensitivity, and 82.00\% \& 88.06\% in F2 score, across ultra-small and small medical object scales, respectively.
Notably, SvANet performance is essentially distinguished in the segmentation of ultra-small tissue cells, surpassing other SOTA models by at least +9.60\% in mIoU, +8.50\% in mDice, and +6.61\% in F2 score.
This superior performance contrasts with improvements of less than 5\% observed in the five other datasets, as shown in \cref{tab:segresults}, which may be attributed to the relatively large number of ultra-small objects in TissuNet (\ie, 9096 cells, approximately ten times more objects than other datasets).

\begin{figure}[!t] %
	\centering
 	\includegraphics[width=\columnwidth]{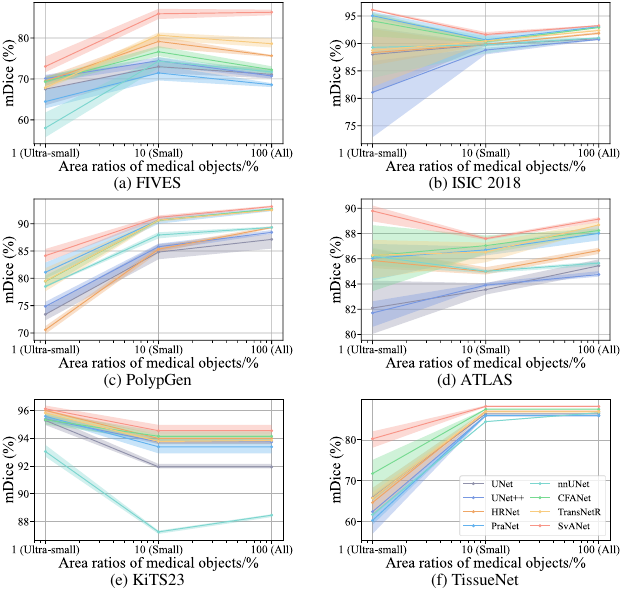} 
 	\caption{Segmentation mDice across different area ratios of medical objects in (a) FIVES, (b) ISIC 2018, (c) PolypGen, (d) ATLAS, (e) KiTS23, and (f) TissueNet datasets. The legends used in these subfigures are consistent with those detailed in the final panel (f).}
 	\label{fig:trends}
\vspace{-1mm}
\end{figure}

Furthermore, the mDice results trends for all tested methods across ultra-small, small, and all medical object segmentation in FIVES, ISIC 2018, PolypGen, ATLAS, KiTS23, and TissueNet datasets are illustrated in \cref{fig:trends}.
This figure highlights the SvANet, represented by the red line, consistently leads across diverse object scales and datasets.
In the FIVES dataset, as shown in \cref{fig:trends}a, only SvANet exhibits an increasing mDice as object scale increases, while other methods' mDice initially increases and then decreases.
The subbranches of retinal vessels are relatively thin, and the number of vessels increases as the occupied area expands.
Therefore, the subbranches become more difficult to discriminate, decreasing mDice as the object scale range expands from $\le$10\% to $\le$100\%.
However, SvANet maintains a growing trend without decline, demonstrating its effectiveness in recognizing retinal vasculatures, which is crucial for diagnosing blindness-causing diseases.
Besides, in ISIC 2018 and KiTS23 datasets, SvANet and over half of other methods exhibit a mDice trend resembling an ``L'' shape, as depicted in \cref{fig:trends}be.
Fewer ultra-small objects in these datasets introduce significant variability, likely contributing to this ``L'' trend.
In the PolypGen, ATLAS, and TissueNet datasets, there is a consistent increase in mDice trends, as shown in \cref{fig:trends}cdf.
Notably, no change is observed in TissueNet between the small and all object scales, as both categories contain identical medical images.
Closer inspection of the \cref{fig:trends}d and \cref{tab:segresults} reveals that SvANet is the only method that achieved a ``V'' trend in the ATLAS dataset, with the best mDice of 89.79\% for segmenting ultra-small compared to small and all sizes of livers and tumors, underscoring SvANet's capability to effectively discriminate ultra-small medical objects.
Additionally, SvANet consistently displays narrow error bars (shown as color bands) across three object scales, indicating its robustness in accurately recognizing medical objects.

\subsection{Results for the Dataset with Only Ultra-Small Objects}\label{sec:results2}
To further evaluate the performance of SvANet in distinguishing ultra-small medical objects, experiments were conducted in the SpermHealth dataset, which exclusively has sperms with an area ratio of less than 1\%.
As shown in \cref{tab:segresults2}, SvANet secured top performance in sperm segmentation within the SpermHealth dataset, achieving 72.58\% in mDice, 61.44\% in mIoU, 13.06\texttimes10\textsuperscript{-4} in MAE, 72.50\% in sensitivity, and 72.51\% F2 score.
SvANet's performance in sperm segmentation notably exceeded that of other models, surpassing them by up to 15.88\% in F2 score, 14.99\% in sensitivity, 14.64\% in mDice, and 11.86\% in mIoU.
Besides, the performance metrics (mDice, mIoU, sensitivity, and F2 score) gained in the SpermHealth dataset are significantly lower than those observed in ISIC 2018, PolypGen, ATLAS, and KiTS23 for all tested models, with a gap of $>$10\%, because all sperms have an area lower than 1\%, presenting limited learnable features and posing more significant challenges for differentiation.

\begin{table}[!t]
	\centering
	\scriptsize
	\renewcommand{\arraystretch}{1.1}
	\setlength\tabcolsep{5pt}
	\caption{
Quantitative results in the SpermHealth dataset. 
All sperms in this dataset occupy less than 1\% of images' area. 
The best results are underlined in bold.
}\label{tab:segresults2}
	\resizebox{\columnwidth}{!}{
		\begin{tabular}{cr|*{5}{|C{0.9cm}}}
	
			\Xhline{4\arrayrulewidth}
			\rowcolor{mygray}
			& Methods & mDice & mIoU & MAE\textsuperscript{*} & Sensitivity & F2 score \\
			\Xhline{2\arrayrulewidth}
		
			\multirow{8}{*}{\begin{sideways}SpermHealth\end{sideways}} 
			& UNet (MICCAI'15)~\cite{ronneberger2015unet} & 58.47 & 50.18 & 13.54 & 57.52 & 57.65 \\
			& UNet++ (TMI'19)~\cite{zhou2019unet++} & 57.94 & 49.58 & 15.09 & 56.15 & 56.63 \\
			& PraNet (MICCAI'20)~\cite{fan2020pranet} & 60.56 & 50.82 & 16.59 & 57.51 & 58.61 \\
			& HRNet (TPAMI'20)~\cite{wang2020deep} & 64.25 & 54.01 & 14.65 & 62.68 & 63.23 \\
			& nnUNet (NM'21)~\cite{isensee2021nnu} & 65.74 & 55.28 & 13.43 & 67.51 & 66.77 \\
			& CFANet (PR'23)~\cite{zhou2023cross} & 60.58 & 50.88 & 20.01 & 59.13 & 59.63 \\
			& TransNetR (MIDL'24)~\cite{jha2024transnetr} & 70.28 & 59.19 & 14.66 & 69.70 & 69.89 \\
			& \textbf{SvANet~(Ours)} & \textbf{72.58} & \textbf{61.44} & \textbf{13.06} & \textbf{72.50} & \textbf{72.51} \\

			\Xhline{4\arrayrulewidth}
	\multicolumn{7}{r}{\textsuperscript{*}\footnotesize{MAE unit: \texttimes10\textsuperscript{-4}}}
	\end{tabular}
}
\vspace{-1mm}
\end{table}

\begin{figure}[!t]
	\centering
 	\includegraphics[width=0.9\columnwidth]{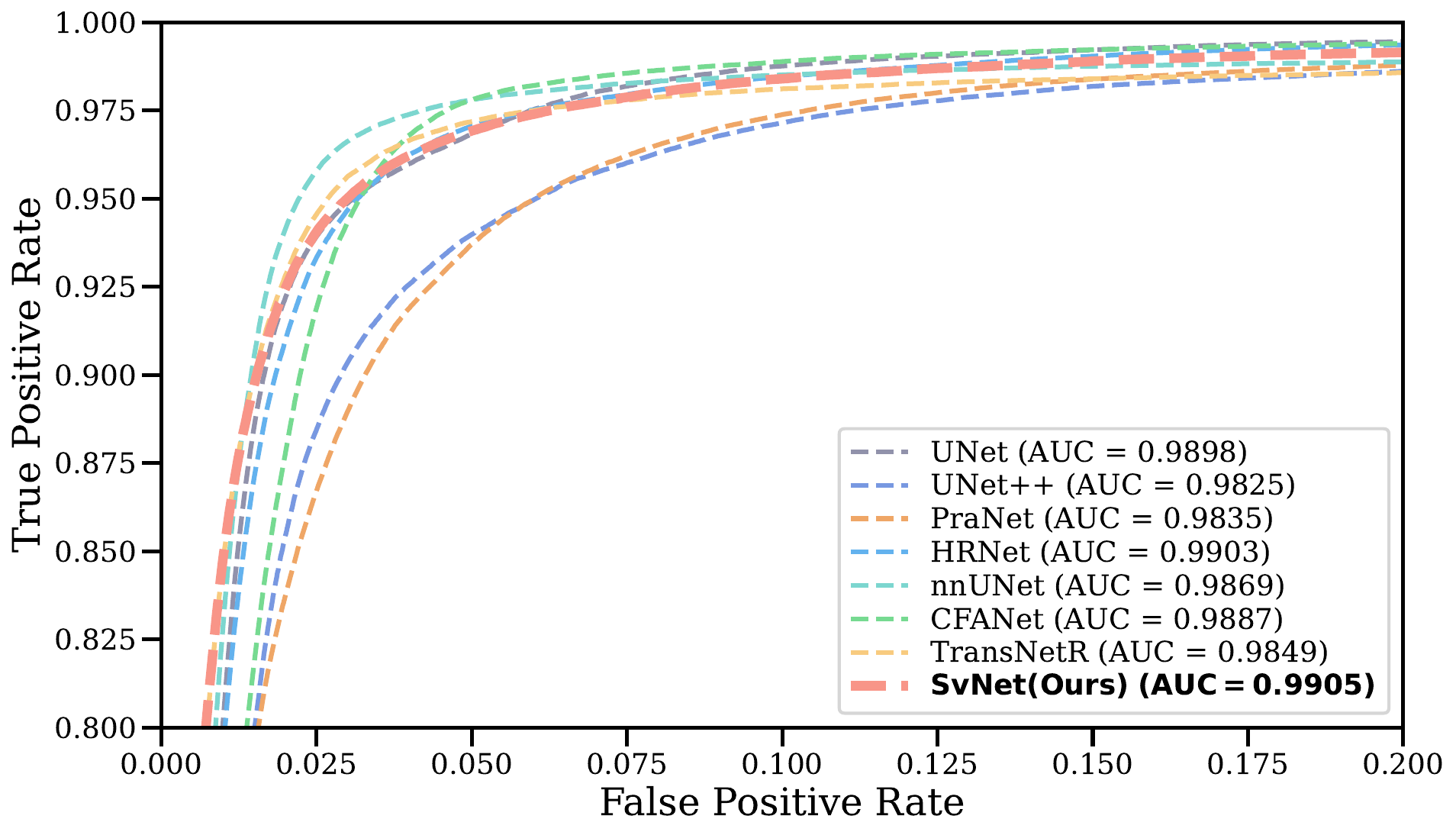}
 	\caption{Receiver operating characteristic (ROC) curves for tested models in the SpermHealth dataset.}
 	\label{fig:rocauc}
\vspace{-1mm}
\end{figure}

To quantify the robustness and adaptability of SvANet vs. other SOTA methods, receiver operating characteristic (ROC) curves of tested methods in the SpermHealth dataset are employed and illustrated in \cref{fig:rocauc}.
The ROC curve of SvANet, represented by the red line in \cref{fig:rocauc}, blends nearest towards the top-left corner, with the highest area under the curve (AUC) of 0.9905, surpassing other SOTA methods by up to AUC of +0.008.
Besides, the ROC curves of UNet++ is close to the lower-right corner and under all other curves, with the lowest AUC of 0.9825.
The ROC and AUC results of UNet++ are consistent with \cref{tab:segresults2}, demonstrating that UNet++ struggled to recognize sperms.

\subsection{Ablation Studies}\label{sec:ablation}
Unless otherwise specified, all ablation studies were conducted in the SpermHealth dataset for the sake of simplicity.
\subsubsection{Main Components Ablation}
To investigate the influence of each core module of SvANet (\ie, MCBottleneck, MCAttn, cross-scale guidance, SvAttn, and AssemFormer), ablation studies were conducted and discussed in this section.
As indicated in the first and second columns of \cref{tab:ablation}, the inclusion of MCBottleneck and MCAttn results in improvements of +0.15\% \& +0.52\% mDice and +0.40\% \& +0.22\% sensitivity, respectively, suggesting that both modules significantly contribute to enhancing the diagnosis and accuracy of positive cases.
Furthermore, when equipped with cross-scale guidance and SvAttn, SvANet achieves an additional +0.41\% \& +0.24\% in mDice and +0.53\% \& +0.51\% in sensitivity.
The ablation of AssemFormer leads to an increase of +0.16\% in mDice and +0.67\% in sensitivity.
The p-value for mDice is below 0.05, confirming the result's reliability.
In summary, MCAttn and cross-scale guidance are critical modules for enhancing prediction accuracy.
At the same time, AssemFormer and MCBottleneck are vital in improving the accuracy of positive diagnostics in medical object segmentation.

\begin{table}[!t]
	\centering
	\scriptsize
	\renewcommand{\arraystretch}{1.1}
	\setlength\tabcolsep{5pt}
	\caption{
Ablation studies results on the main components of SvANet.
The best results are underlined in bold.
}\label{tab:ablation}
	\resizebox{\columnwidth}{!}{
		\begin{tabular}{C{1.3cm}*{4}{|C{1.3cm}}*{3}{|C{1.0cm}}}
	
			\Xhline{4\arrayrulewidth}
			\rowcolor{mygray}
			\multicolumn{5}{c|}{Ablation settings} & & & \\ [-0.5pt] \hhline{*{5}{-}}
			\rowcolor{mygray}
			& & Cross-scale & & & & & \\
			\rowcolor{mygray}
			\multirow{-2}{*}{MCBottleneck} & \multirow{-2}{*}{MCAttn} & guidance & \multirow{-2}{*}{SvAttn} & \multirow{-2}{*}{AssemFormer} & \multirow{-3}{*}{mDice} & \multirow{-3}{*}{Sensitivity} &  \multirow{-3}{*}{p-value} \\
			\Xhline{2\arrayrulewidth}

			&  &  &  &  & 71.10 & 70.08 & - \\
			\cmark &  &  &  &  & 71.25 & 70.48 & 0.042 \\
			\cmark & \cmark &  &  &  & 71.77 & 70.70 & 0.033 \\
			\cmark & \cmark & \cmark &  &  &  72.18 & 71.33 & 0.019 \\
			\cmark & \cmark & \cmark & \cmark &  & 72.42 & 71.84 & 0.013 \\
			\cmark & \cmark & \cmark & \cmark  & \cmark & \textbf{72.58} & \textbf{72.51} & \textbf{0.001} \\

			\Xhline{4\arrayrulewidth}
	\end{tabular}
}
\vspace{-1mm} 
\end{table}

\subsubsection{MCAttn vs. Other Advanced Attention Methods}
To assess the impact of different attention mechanisms within the MCBottleneck, three advanced attention modules, including squeeze-excitation (SE), convolution block attention module (CBAM), and coordinate attention (CoorAttn), were utilized as the control group.
According to the results shown in \cref{tab:attnmethod}, MCAttn achieved performance improvements of over +1.15\% in mDice and +1.12\% sensitivity compared to these alternatives. 
Notably, the control group's attention methods resulted in reduced performance, with decreases of up to -0.83\% in mDice and -0.41\% in sensitivity, underscoring the superior efficacy of MCAttn in enhancing medical image segmentation within a bottleneck structure.
The p-value for mDice is below 0.05, affirming the reliability of the result.

\begin{table}[!t]
	\centering
	\scriptsize
	\renewcommand{\arraystretch}{1.2}
	\setlength\tabcolsep{5pt}
	\fontsize{7pt}{7pt}\selectfont
	\caption{Comparison of MCAttn with other advanced attention methods. 
The best results are underlined in bold.}
	\label{tab:attnmethod}
	\resizebox{\columnwidth}{!}{
		\begin{tabular}{r*{4}{|C{1.4cm}}}
			
			\Xhline{4\arrayrulewidth}
			\rowcolor{mygray}
			 & \# Parameters &  &  & \\
			\rowcolor{mygray}
			\multirow{-2}{*}{Attention module} & /Million & \multirow{-2}{*}{mDice} & \multirow{-2}{*}{Sensitivity} & \multirow{-2}{*}{p-value} \\
			\Xhline{2\arrayrulewidth}

			- & - & 71.81 & 70.37 & - \\ 
			SE~\cite{hu2018squeeze} & 2.77 & 71.19 & 71.39 & 0.048 \\ 
			CBAM~\cite{woo2018cbam} & 0.70 & 70.98 & 69.96 & 0.041 \\ 
			CoorAttn~\cite{hou2021coordinate} & 0.10 & 71.43 & 70.14 & 0.049\\ 
			\textbf{MCAttn (ours)} & 2.77 & \textbf{72.58} & \textbf{72.51} & \textbf{0.022}\\ 

			\Xhline{4\arrayrulewidth}
	\end{tabular}
}
\vspace{-1mm}
\end{table}	

\subsubsection{The Number of Pooled Tensors for MCAttn}
The selection of the size and number of pooled tensors for MCAttn is crucial for expanding network variants.
We tested combinations (1, 2), (1, 2, 3), (2, 3), and (1, 2, 3, 4).
The results, shown in the first, second, and fourth rows of \cref{tab:poolnum}, reveal that the (1, 2, 3) combination of pooled tensors outperformed (1, 2) and (1, 2, 3, 4) combinations, with improvements exceeding 2.86\% \& 1.19\% in mDice and 4.01\% \& 1.66\%, respectively.
Further analysis, as indicated in the second and third rows of \cref{tab:poolnum}, highlights the necessity of a pool size of 1, leading to an increase of 0.82\% in mDice and +2.27\% in sensitivity.
These findings emphasize the importance of maintaining an optimal level of variation in the network.
An insufficient pooled tensor can limit performance, whereas an excessive number can introduce too much stochasticity.
Thus, striking the right balance is critical for maximizing the effectiveness of MCAttn within the model.

\begin{table}[!t]
	\centering
	\scriptsize
	\renewcommand{\arraystretch}{1.2}
	\setlength\tabcolsep{5pt}
	\caption{Size combinations of pooled feature maps in MCAttn (Unit: \%). 
The best results are underlined in bold.}
	\label{tab:poolnum}
	\resizebox{0.68\columnwidth}{!}{
		\begin{tabular}{r*{2}{|C{1.3cm}}}
	
			\Xhline{4\arrayrulewidth}
			\rowcolor{mygray}
			Pooled tensor sizes & mDice & Sensitivity \\
			\Xhline{2\arrayrulewidth}

			(1, 2) & 69.72 & 68.50 \\
			(\textbf{1, 2, 3}) & \textbf{72.58} & \textbf{72.51} \\
			(2, 3) &  71.76 & 70.24 \\
			(1, 2, 3, 4) & 71.39 & 70.85 \\
			
			\Xhline{4\arrayrulewidth}
	\end{tabular}
}
\vspace{-1mm}
\end{table}

\subsubsection{Inference Time}
This section quantifies the inference characteristics of the tested networks, including the number of parameters (\# Parameters), multiply-accumulate operations (MACs), and inferencing speed.
The unit of inference speed is frames per second (FPS).
The number of classes was set to eight, and other configurations were consistent with those described in \cref{sec:imple}.
The inference speed results, averaged over 1000 runs, are presented in \cref{tab:inference}.

\begin{table}[!t]
	\centering
	\scriptsize
	\caption{Model size and inference time comparison.}
	\label{tab:inference}
	\begin{tblr}{
	vline{2-Y} = {},
	hline{1,Z} = {4\arrayrulewidth},
	hline{3} = {2\arrayrulewidth},
	row{1,2} = {bg=mygray},
	column{2-Z} = {wd=13mm},
	cells = {c,m},
	cell{1}{1} = {r=2}{halign=r},
	cell{2-Z}{1} = {halign=r},
}
			Methods & \# Parameters & MACs & Speed \\
			& /Million & /Billion & /FPS \\

			UNet (MICCAI'15)~\cite{ronneberger2015unet} & 34.53 & 262.21 & 93 \\
			UNet++ (TMI'19)~\cite{zhou2019unet++} & 9.16 & 139.68 & 100 \\
			PraNet (MICCAI'20)~\cite{fan2020pranet} & 30.34 & 25.65 & 82 \\
			HRNet (TPAMI'20)~\cite{wang2020deep} & 63.60 & 65.80 & 78 \\
			nnUNet (NM'21)~\cite{isensee2021nnu}& 30.60 & 232.19 & 83 \\
			CFANet (PR'23)~\cite{zhou2023cross} & 25.71 & 115.63 & 56 \\
			TransNetR (MIDL'24)~\cite{jha2024transnetr} & 27.27 & 44.71 & 100 \\
			\textbf{LiteSvANet~(Ours)} & 53.04 & 224.18 & 77 \\ 
			\textbf{SvANet~(Ours)} & 177.64 & 312.76 & 55 \\
	\end{tblr}
\end{table}

\cref{tab:inference} illustrates that SvANet achieved a real-time analysis of medical images with 55 FPS.
Notably, while SvANet consumed 312.76 billion MACs --- two times more than CFANet --- it performed at only 1 FPS lower than CFANet.
This discrepancy highlights that MACs, as theoretical indicators of computational cost, do not fully capture the effects of hardware or software optimizations for inferencing.
Despite the high computational load, SvANet's performance remains well-suited for self-examination and clinical diagnostic applications.

Additionally, a streamlined version of SvANet, named LiteSvANet, was developed by omitting the fifth encoder stage while retaining the ASPP on the fourth encoder stage, reducing the parameter count by 70\%.
Subsequent tests, under identical conditions (described in \cref{sec:imple}), demonstrated that LiteSvANet achieved a mDice of 70.88\% and a sensitivity of 69.96\% in the SpermHealth dataset, surpassing the performance of the second-best method, TransNetR, as shown in \cref{tab:segresults2}.
Moreover, LiteSvANet significantly enhanced the inference speed to 77 FPS, which, while 23\% lower than the fastest models (UNet ++ and TransNetR), represents a considerable improvement over the standard SvANet model.
For straightforward applications, implementing LiteSvANet is advantageous for examining small medical objects.

\subsection{Negative Case Studies}\label{sec:vis}
\begin{figure*}[!tb]
	\centering
 	\includegraphics[width=\textwidth]{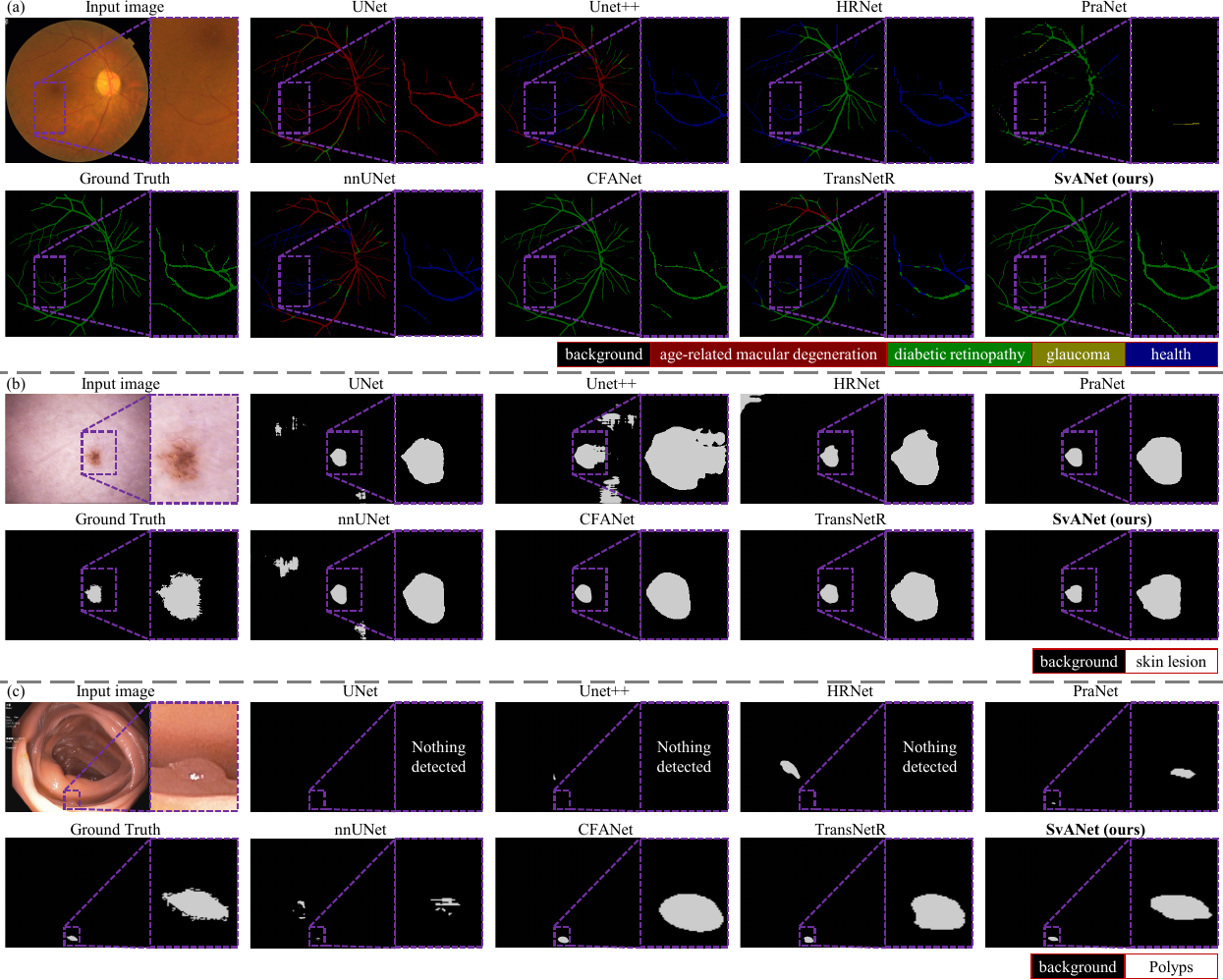}
 	\caption{
Examples of segmentation results across tested methods in the (a) FIVES, (b) ISIC 2018, and (c) PolypGen datasets for error analysis. 
Examples contain ultra-small objects (\ie, polyp), small objects (\ie, nevus), and objects with $>$10\% area ratio (\ie, retinal vessel).
}
 	\label{fig:vis1}
\vspace{-1mm}
\end{figure*}

\begin{figure*}[!tb]
	\centering
 	\includegraphics[width=\textwidth]{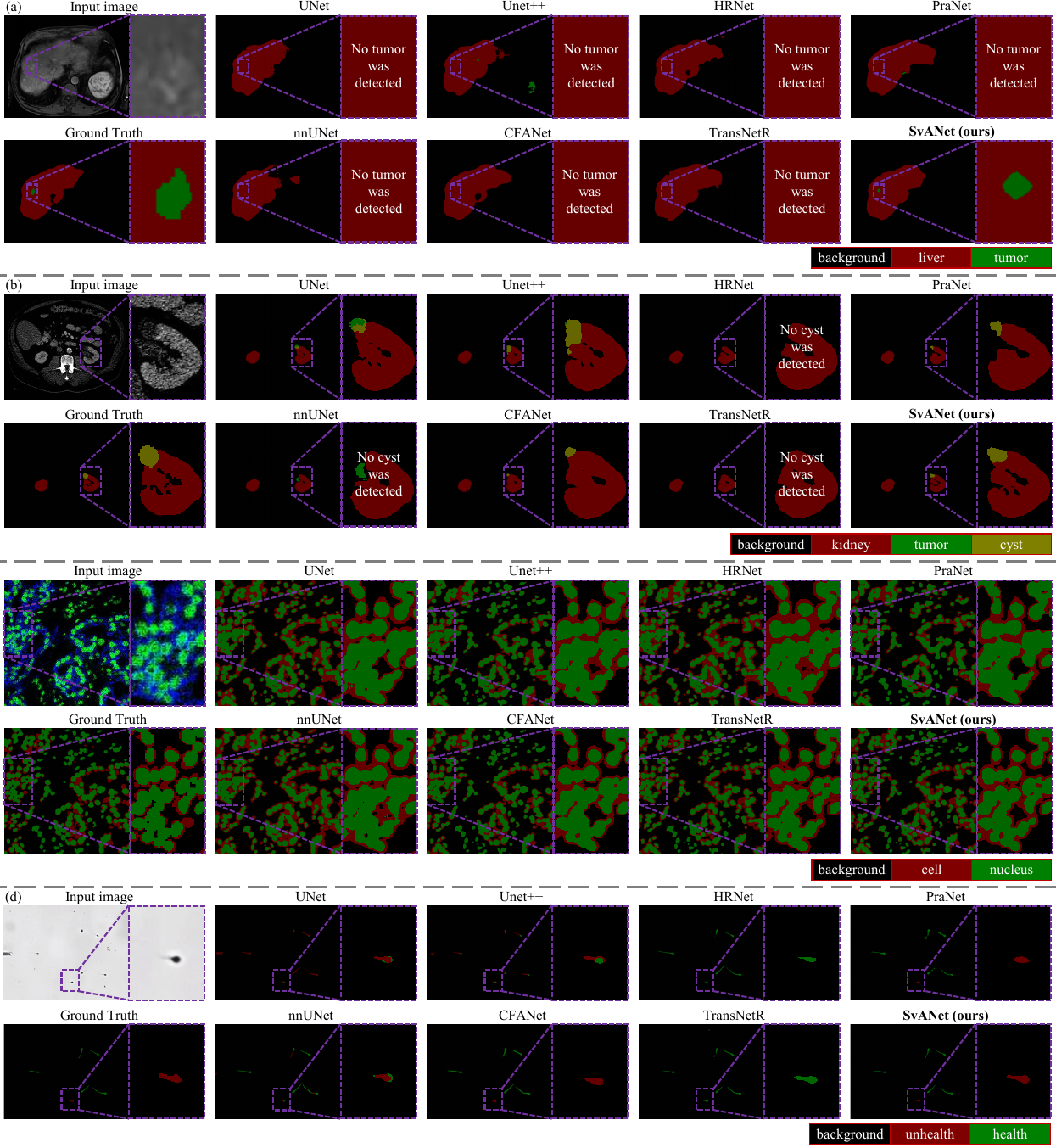}
 	\caption{Examples of segmentation results across tested methods in the (a) ATLAS, (b) KiTS23, (c) TissueNet, and (d) SpermHealth datasets for error analysis. 
Examples contain ultra-small objects (\ie, tumor, cyst, tissue cell, nucleus, and sperm), small objects (\ie, kidney), and objects with $>$10\% area ratio (\ie, liver).}
 	\label{fig:vis2}
\vspace{-1mm}
\end{figure*}

Examples of visualization results for ultra-small and small medical objects in the FIVES, PolypGen, ISIC 2018, ATLAS, KiTS23, TissueNet, and SpermHealth datasets are presented in \cref{fig:vis1} and \cref{fig:vis2}.
As illustrated in \cref{fig:vis1}a, UNet misclassified diabetic retinopathy (green region) as age-related macular degeneration (red region).
Similarly, UNet++, HRNet, nnUNet, and TransNetR misclassified diabetic retinopathy as health retinal vessels (blue region), while PraNet misclassified it as glaucoma (yellow region).
Additionally, PraNet struggled to detect retinal vasculature in the zoomed-in region.
Furthermore, none of the SOTA methods in the control group could accurately recover the retinal vessels at the bottom of the zoomed-in region.
In contrast, SvANet not only correctly classified diabetic retinopathy but also effectively detected the position and shape of retinal vessels.

In the skin lesion examination illustrated in \cref{fig:vis1}b, UNet, UNet++, HRNet, nnUNet misclassified normal skin as a nevus.
Moreover, UNet, PraNet, nnUNet, CFANet, and TransNetR represented the nevus region as a relatively smooth circle, while UNet++ and HRNet captured a larger region encompassing the ground truth annotations, leading to an underestimation of the lesion boundary.
In contrast, SvANet accurately identified a skin lesion of similar size to the ground truth and delineated its sawtooth-shaped boundary.
For the polyps diagnosis as presented in \cref{fig:vis1}c, SOTA methods such as PraNet and nnUNet either detected a smaller polyp area than the ground truth, or other methods, including CFANet and TransNetR, regarded a larger region than the ground truth.
Besides, UNet, UNet++, and HRNet failed to detect the polyp in the example image.
Furthermore, the detected regions from methods in the control group significantly deviated from the ground truth.
However, SvANet recognized an area close to the ground truth and maintained shapes akin to ground truth annotations.

For MRI and CT image modalities analysis, as shown in \cref{fig:vis2}ab, it is possible to overlook the overlapping medical objects, particularly ultra-small ones.
For instance, all tested models in the control group failed to identify an ultra-small tumor inside the liver.
In addition, HRNet, nnUNet, and TransNetR missed an ultra-small cyst at the edge of the kidney.
Moreover, UNet++ and CFANet incorrectly emphasized the background as a tumor or liver region in the example image, and UNet misclassified a cyst as a tumor.
Although the organ region (\eg, liver and kidney) detected by SOTA methods in the control group appeared complete, the pathological regions, such as the tumor and cyst, were either larger (hepatic tumor) or smaller (cyst) than the ground truth in the example image.
However, SvANet accurately differentiated between organs and their pathological regions.
Furthermore, SvANet captured the morphological details of the liver, hepatic tumor, kidney, and cyst in the example image, closely aligning with the ground truth annotations.

For tissue cell recognition in the TissueNet dataset, as shown in \cref{fig:vis2}c, both TransNetR and SvANet effectively delineated cell boundaries and accurately labeled the cells and nuclei regions, closely resembling the ground truth. 
In contrast, other SOTA methods struggled to categorize cells and nuclei, leading to difficulties in differentiating cell boundaries and merging several cells.
For sperm cell analysis, as presented by the final image in \cref{fig:vis2}d, SvANet precisely located all sperm positions and effectively recognized the region of the short tail of an abnormal sperm.
Conversely, tested methods like PraNet and CFANet struggled to differentiate the head and tail of the unhealthy sperm, as illustrated in the zoomed-in region of \cref{fig:vis2}d.
Moreover, UNet, UNet++, HRNet, nnUNet, and TransNetR misclassified an unhealthy sperm head as healthy, as indicated by a green subregion in \cref{fig:vis2}d.

These visualization results align with the findings discussed in \cref{sec:results1} and \cref{sec:results2}, suggesting that SvANet holds significant potential for application in general small medical object recognition across various medical imaging modalities for disease diagnostics, surgeries, \etc.

\section{Conclusion}\label{sec:conclu}
This paper introduces SvANet, a novel network to enhance the segmentation of small medical objects, aiding in the detection of life-threatening diseases and supporting in vitro fertilization.
The experimental results demonstrate that the SvANet is significantly effective in distinguishing medical objects of various sizes.
SvANet consistently outperformed other SOTA methods, achieving up to 19.95\%, 15.03\%, 15.01\%, 14.64\%, 13.57\%, 8.09\%, and 3.07\% increments in mDice for segmenting objects occupying less than 1\% image area across TissueNet, FIVES, ISIC 2018, SpermHealth, PolypGen, ATLAS, and KiTS23 datasets.
Furthermore, the visualization results confirm that SvANet accurately identifies the locations and morphologies of all medical objects, demonstrating its exceptional capability in segmenting small medical objects.
These findings underscore the potential of SvANet as a significant advancement in medical imaging.

\bibliographystyle{IEEEtran}
\bibliography{egbib}

\clearpage
\section*{Supplemental Materials}

This section discusses auxiliary but noteworthy settings and circumstances related to this work.
Additional experiments presented herein aim to clearly demonstrate the advancements offered by our models.

\subsection*{Area Computation}
The same instance across different images is treated as an independent object because several datasets do not have instance labels during inference.
For a fair comparison, each object instance is counted once per image.
Area computation results indicate that there are 18\% of retinal vasculatures in FIVES~\cite{jin2022fives}, 40\% of skin lesions in ISIC 2018~\cite{codella2019skin}, 63\% of polyps in PolypGen~\cite{ali2023multi}, 74\% of hepatic tumors and livers in ATLAS~\cite{quinton2023tumour}, 99\% of kidneys and corresponding neopasms in KiTS23~\cite{heller2023kits21}, and 100\% tissue cells in TissueNet~\cite{greenwald2022whole} under 10\% area ratio.
The area calculation codes are publically available at \url{https://github.com/anthonyweidai/circle_extractor}.

\subsection*{Network Instantiation}\label{sec:instant}
The input medical image is initially processed through a stem module, which consists of a strided convolution that reduces the resolutions of the image by half. 
The output resolutions at the end of the five stages are 1/2, 1/4, 1/8, 1/16, and 1/32.
The output channels of the bottom row blocks are 64, 64, 128, 256, and 512, respectively.

\subsection*{Extracurricular Ablation}
The configurations of training and testing are the same as that in \cref{sec:ablation} unless specified.

\subsubsection*{Place of MCAttn}
The ablation study presented in \cref{tab:mcplace} demonstrates that using MCAttn exclusively within MCBottleneck delivered at least +2.04\% in mDice and +3.06\% in sensitivity compared to its application in AssemFormer.

\begin{table}[!h]
    \centering
    \caption{
Impact of the attention modules in MCBottleNeck and AssemFormer in the SpermHealth dataset. 
Configurations include (0): None, (1): Squeeze-excitation, and (2) MCAttn. 
The best results are underlined in bold.
}
    \label{tab:mcplace}
    \resizebox{0.8\columnwidth}{!}{
        \begin{tabular}{C{1.6cm}|C{1.6cm}*{2}{|C{1.0cm}}}
            \Xhline{4\arrayrulewidth}

			\rowcolor{mygray}
			Attention in & Attention in & & \\
			\rowcolor{mygray}
			MCBottleneck & AssemFormer & \multirow{-2}{*}{mDice} & \multirow{-2}{*}{Sensitivity}\\
			\Xhline{2\arrayrulewidth}

			(0) & (0) & 71.81 & 70.37 \\
			(1) & (1) & 69.99 & 68.19 \\
			(2) & (2) & 70.54 & 69.45 \\
			(2) & (0) & \textbf{72.58} & \textbf{72.51} \\
			\Xhline{4\arrayrulewidth}
        \end{tabular}
    }
\vspace{-0mm}
\end{table}

\subsubsection*{Place of AssemFormer}
Given that AssemFormer operates before and after the concatenation operation, exploring its impact in these settings is essential.
As indicated in \cref{tab:afplace}, positioning AssemFormer only before or after concatenation leads to reductions in performance, with drops ranging from 1.25\% $\sim$ 1.30\% in mDice and 1.60\% $\sim$ 1.86\% in sensitivity.
Conversely, employing AssemFormer in both locations yields improvements of 0.16\% in mDice and 0.67\% in sensitivity, enhancing sperms diagnosis.
Such results suggest the importance of consistent tensor operations before and after concatenation for cross-scale guidance, as depicted in \cref{fig:archi}, ensuring that tensors are processed in the same latent space.

\begin{table}[!h]
	\centering
	\scriptsize
	\renewcommand{\arraystretch}{1.2}
	\setlength\tabcolsep{5pt}
	\caption{Impact of the AssemFormer module located before and after concatenation in the SpermHealth dataset. 
The best results are underlined in bold.}
	\label{tab:afplace}
	\resizebox{0.9\columnwidth}{!}{
		\begin{tabular}{C{2cm}|C{2cm}*{2}{|C{1.3cm}}}
			
			\Xhline{4\arrayrulewidth}
			\rowcolor{mygray}
			\multicolumn{2}{c|}{AssemFormer} & & \\ [-0.5pt] \hhline{*{2}{-}}
			\rowcolor{mygray}
			Before & After &  & \\
			\rowcolor{mygray}
			concatenation & concatenation & \multirow{-3}{*}{mDice} & \multirow{-3}{*}{Sensitivity}\\
			\Xhline{2\arrayrulewidth}

			&  & 72.42 & 71.84 \\
			\cmark &  & 71.17 & 69.98 \\
			& \cmark & 71.12 & 70.34 \\
			\cmark  & \cmark & \textbf{72.58} & \textbf{72.51} \\

			\Xhline{4\arrayrulewidth}
	\end{tabular}
}
\vspace{-0mm}
\end{table}

\subsubsection*{Concatenation or Addition after MCBottleneck}
As shown in \cref{fig:archi}, post-MCBottleneck concatenation operations can be replaced using direct links or addition operations.
Nevertheless, the ablation results detailed in \cref{tab:concatenate} demonstrate that concatenation for post-MCBottleneck is more effective.
Specifically, concatenation delivered improvements of 3.42\% \& 1.43\%  in mDice and 6.07\% \& 3.12\% in sensitivity, compared to direct links and addition operations, separately.
The findings underscore the importance of preserving feature characteristics and depth information in outputs from the MCBottleneck and cross-scale guidance.

\begin{table}[!h]
	\centering
	\scriptsize
	\renewcommand{\arraystretch}{1.2}
	\setlength\tabcolsep{5pt}
	\caption{
Comparison of concatenation versus addition operations after MCBottleneck in the SpermHealth dataset. 
The best results are underlined in bold.
}
	\label{tab:concatenate}
	\resizebox{0.8\columnwidth}{!}{
		\begin{tabular}{r*{3}{|C{1.3cm}}}
			
			\Xhline{4\arrayrulewidth}
			\rowcolor{mygray}
			 & \# Parameters &  & \\
			\rowcolor{mygray}
			\multirow{-2}{*}{Settings} & /Million & \multirow{-2}{*}{mDice} & \multirow{-2}{*}{Sensitivity} \\
			\Xhline{2\arrayrulewidth}

			- & - & 69.16 & 66.44 \\ 
			Addition & 12.17 & 71.15 & 69.39 \\ 
			\textbf{Cancatenation} & 38.41 & \textbf{72.58} & \textbf{72.51} \\ 

			\Xhline{4\arrayrulewidth}
	\end{tabular}
}
\vspace{-0mm}
\end{table}


%

\end{document}